\documentclass[aps,prb,floatfix,twocolumn,showpacs,superscriptaddress]{revtex4-2}
\usepackage{graphicx}
\usepackage{amsmath,amssymb}
\usepackage{flushend}
\usepackage[utf8]{inputenc}
\usepackage[colorlinks = true,
linkcolor = blue,
urlcolor  = blue,
citecolor = blue,
anchorcolor = blue]{hyperref}
\usepackage{array}
\usepackage{bbold}
\usepackage{float}
\usepackage{enumitem}
\usepackage{braket}
\usepackage[mathscr]{euscript}
\newcommand{\RNum}[1]{\uppercase\expandafter{\romannumeral #1\relax}}
\usepackage{booktabs}
\mathchardef\Re="023C
\mathchardef\Im="023D
\usepackage{color}
\usepackage{times}
\usepackage{setspace}
\begin{document}
	
\title{Parsing skin effect in a non-Hermitian spinless BHZ-like model}
	
\author{Dipendu Halder}\email[corresponding author: ]{h.dipendu@iitg.ac.in}
\author{Saurabh Basu}\email[]{saurabh@iitg.ac.in}
\affiliation{Department of Physics, Indian Institute of Technology Guwahati-Guwahati, 781039 Assam, India}
	
\begin{abstract}
This work comprehensively investigates the non-Hermitian skin effect (NHSE) in a spinless Bernevig-Hughes-Zhang (BHZ)-like model in one dimension. It is generally believed that a system with non-reciprocal hopping amplitudes demonstrates NHSE. However, we show that there are exceptions, and more in-depth analyses are required to decode the presence of NHSE or its variants in a system. The fascinating aspects of our findings, depending on the inclusion of non-reciprocity in the inter-orbital hopping terms, concede the existence of conventional NHSE or NHSE at both edges and even a surprising absence of NHSE. The topological properties and the (bi-orthogonal) bulk-boundary correspondence, enumerated via computation of the (complex) Berry phase and spatial localization of the edge modes, highlight the topological phase transitions occurring therein. Further, to facilitate a structured discussion of the non-Hermitian model, we split the results into $\mathcal{PT}$ symmetric and non-$\mathcal{PT}$ symmetric cases with a view to comparing the two.		
\end{abstract}

\maketitle
 

\section{\label{sec1}Introduction}

The Hermiticity of a Hamiltonian for a quantum mechanical system ensures the energy is real and a conserved quantity.
On the contrary, open quantum systems can demonstrate non-Hermitian (NH) dynamics, resulting in markedly distinct behavior from those for the Hermitian systems.
Ever since the pioneering discovery \cite{PhysRevLett.80.5243} by Bender and Boettcher of NH systems with $\mathcal{PT}$ (product of parity and time-reversal symmetries) symmetry exhibiting real energy spectra, these systems have become intriguing subjects for further investigations.
Along with the mathematical advances in NH physics \cite{Bender1999, PhysRevLett.89.270401, Mostafazadeh_2004, moiseyev_2011}, recent experimental studies on NH properties in optical systems \cite{ Eichelkraut, Xiao, Weimann, PhysRevLett.123.230401, Wang}, electronic systems \cite{ PhysRevLett.123.193901, Helbig2020, PhysRevLett.126.215302}, acoustic systems \cite{Fleury2015, PhysRevApplied.16.057001, Zhang2021} have prompted the exploration of this field in a significant way.

Further, applications of topology in condensed matter \cite{RevModPhys.82.3045, RevModPhys.83.1057, RevModPhys.88.035005} physics have started playing a major role in understanding the properties of materials since the discovery of the quantum Hall effect \cite{PhysRevLett.49.405}.
These are known as topological insulators (TIs) and are at the center stage of exploration in recent times.
They obey certain symmetries, and their topological properties are characterized by different symmetry classes they fall in \cite{PhysRevB.55.1142, PhysRevB.78.195125, Ryu_2010}.
The influence of topology in tight-binding models \cite{PhysRevLett.95.226801, PhysRevLett.98.106803, PhysRevLett.42.1698, PhysRevB.22.2099, PhysRevLett.61.2015, PhysRevLett.95.146802} has been being scrutinized for the past few decades.
An example of such a model is the Bernevig-Hughes-Zhang (BHZ) model \cite{BHZ2006}, renowned for exhibiting the quantum spin Hall effect (QSHE) in HgTe/CdTe quantum well structures.
The quantum spin Hall state can be conceptualized as two replicas of the quantum Hall state, each representing one spin orientation.
The transition from a trivial insulator to a TI is observed via a band inversion beyond a certain critical thickness of the HgTe structure.

Here, we explore the topological characteristics of a one-dimensional NH variant of the BHZ-like model, which comprises of spinless fermions that occupy the $s$ and $p_x$ orbitals within a unit cell.
It is important to note that while the original BHZ model is a proposal for a spinful 2D system and is well-known for demonstrating the QSHE, our investigation on the spinless variant in 1D excludes any relevance to QSHE and focuses on its NH properties.
Generally, to study topology in 1D tight-binding models, one can define specific physical quantities, known as topological invariants, such as the winding number \cite{PhysRevB.82.115120}, Berry phase \cite{Berry, M_V_Berry_1985} etc., that differentiate between topologically non-trivial phases and trivial ones.
Specifically, one can investigate the edge states in an open boundary condition (OBC) scenario and compute the Berry phase in periodic boundary conditions (PBC).
The characteristics of these systems undergo substantial modifications when non-Hermiticity is introduced, and show exotic behaviors like breaking down of bulk-boundary correspondence (BBC) \cite{PhysRevLett.116.133903, PhysRevLett.121.026808, PhysRevLett.121.086803, PhysRevB.99.081302, PhysRevB.99.081103, PhysRevB.84.205128}, emergence of exceptional points (EP) \cite{Smilga_2009, Kato, Heiss_2012, PhysRevA.100.062131, Chen_2019, Ozdemir2019, Halder_2023}, complex energy gaps \cite{PhysRevLett.77.570, PhysRevLett.80.5172, PhysRevX.8.031079, Halder_2023}, non-Hermitian skin effect (NHSE) \cite{PhysRevB.97.121401, PhysRevLett.121.086803, YUCE2020126094, PhysRevLett.124.086801, PhysRevB.102.205118, PhysRevLett.124.056802, PhysRevResearch.2.013280, Halder_2023} etc.
These recent studies hint towards the necessity of new fundamental theories for the band topology of the tight binding systems in the NH platform \cite{Ashida2020, PhysRevX.9.041015, Ghatak_2019, PhysRevX.8.031079, RevModPhys.93.015005}.

We introduce non-Hermiticity into the Hermitian model through two distinct methods: introducing a staggered onsite imaginary potential and incorporating non-reciprocal hopping amplitudes.
Hence, we classify such NH systems based on their adherence to the $\mathcal{PT}$ symmetry.
Furthermore, we observe the existence of bi-orthogonal BBC (BBBC) \cite{PhysRevLett.121.026808} within a distinct non-reciprocal model, suggesting that NHSE does not occur in this specific system.
To explore the topological properties of the model, we proceed to calculate the topological invariant, which is the complex Berry phase (CBP) \cite{GARRISON1988177, G_Dattoli_1990, MOSTAFAZADEH199911}, and the inverse participation ratio (IPR) to observe the behavior of the edge modes.
Additionally, we identify a different kind of non-reciprocal system that exhibits both conventional NHSE and NHSE at both edges, with only a slight alteration in the non-reciprocity parameter.
Additionally, we pinpoint the EPs and the conditions in which they emerge within the system.

Our paper is organized as follows.
In section \ref{sec2}, we provide a concise introduction to the Hermitian model and its representations in real and momentum space.
This will assume a smooth comparison with different NH variants that we shall be discussing later.
This is followed by a systematic presentation of the results obtained from the Hermitian model.
Having analyzed the behavior of the edge modes via calculating the IPR and the energy spectra, we compute the topological invariant, which in this case is the Berry phase.
In section \ref{sec3}, we introduce the two non-$\mathcal{PT}$ symmetric NH models, categorizing them based on the presence and absence of the time-reversal symmetry (TRS).
For the models, obeying the BBBC, it is evident that the topological invariant, namely CBP, takes on values $0$ or $-\pi$, corresponding to the trivial and topological phases, respectively.
In section \ref{sec4}, we delve into the discussion of the $\mathcal{PT}$ symmetric NH model, characterized by non-reciprocal hopping amplitudes.
The (non-)existence of NHSE is carefully scrutinized in each case.
In section \ref{sec5}, we summarize the results.

\section{\label{sec2}Model description}
We begin by discussing a 1D extension of the BHZ-like model comprising of $s$ and $p_x$ orbitals in a unit cell.
There are no orbital angular momentum quantum numbers associated with the orbitals, that is, $l=0$ ($s$ orbital) or $l=1$ ($p_x$ orbital), however, we retain the nomenclature owing to the similarity of the BHZ model.
The Hamiltonian of the system is given by,
\begin{align}
		H_0=\sum_{i=1}^{L}\bigg[\epsilon_{s}\hat{s}_{i}^{\dagger}\hat{s}_{i}+\epsilon_{p}\hat{p}_{i}^{\dagger}\hat{p}_{i}\bigg]-&\sum_{i=2}^{L}t_{ps}\hat{s}_{i}^{\dagger}\hat{p}_{i-1}+\nonumber\\\sum_{i=1}^{L-1}\bigg[-t_{s}\hat{s}_{i}^{\dagger}\hat{s}_{i+1}+t_{p}&\hat{p}_{i}^{\dagger}\hat{p}_{i+1}+t_{ps}\hat{s}_{i}^{\dagger}\hat{p}_{i+1}\bigg]+\textrm{H.c.},
	\label{eq:Ham1}
\end{align}
where $\epsilon_s$ and $\epsilon_p$ are the onsite potentials corresponding to the $s$ and $p_x$ orbitals, $-t_s$($t_p$) and $t_{ps}$($-t_{ps}$) being the hopping strengths for $s^{i-1}\leftrightarrow s^i$($p_x^{i-1}\leftrightarrow p_x^i$) and $s^{i-1}\leftrightarrow p_x^i$($p_x^{i-1}\leftrightarrow s^i$), respectively (double-headed arrows denote the Hermitian case).
$s^i$ and $p_x^i$ suggest $s$ and $p_x$ orbitals at $i^{\mathrm{th}}$ unit cell respectively and $L$ denotes the total number of unit cells.
$\hat{s}_{i}$ ($\hat{s}_{i}^{\dagger}$) and $\hat{p}_{i}$ ($\hat{p}_{i}^{\dagger}$) are annihilation (creation) operators for spinless fermions pertaining to the $s$ and $p_x$ orbitals of the $i^{\mathrm{th}}$ unit cell respectively.
Under PBC, the Hamiltonian in equation \eqref{eq:Ham1} can be written in the following Bloch form in momentum space as,
	\begin{gather}
		H_0(k)=\sum_{k}
		\begin{pmatrix}
			\hat{s}_k^{\dagger} & \hat{p}_k^{\dagger}
		\end{pmatrix}h_0(k)
		\begin{pmatrix}
			\hat{s}_k\\
			\hat{p}_k
		\end{pmatrix},
		\label{eq:bloch}
	\end{gather}
where $\hat{s}_{k}$ ($\hat{s}_k^{\dagger}$) and $\hat{p}_{k}$ ($\hat{p}_k^{\dagger}$) denote the annihilation (creation) operators in momentum space and $p_{i}$($p_{i}^{\dagger}$) with $h_0(k)$ being the Bloch Hamiltonian which has a form,
	\begin{gather}
		h_0(k)=\begin{pmatrix}
			\epsilon_{s}-t_{s}e^{-ik}-t_{s}^{*}e^{ik} & 2it_{ps}\sin k\\-2it_{sp}\sin k & \epsilon_{p}+t_{p}e^{-ik}+t_{p}^{*}e^{ik}
		\end{pmatrix}.
		\label{eq:kspace_1}
	\end{gather}

Before going to the NH versions of the system described by equation \eqref{eq:Ham1}, first, we need to understand the properties of the Hermitian counterpart.
For this, we shall study a special case of this Hamiltonian by setting $\epsilon_{s}=-\epsilon_{p}=\epsilon$ and $t_s=t_p=t_s^*=t_p^*=t$.
With this simplification, the model Hamiltonian in real space can be written as,
\begin{align}
		H_1&=\sum_{i=1}^{L}\bigg[\epsilon(\hat{s}_{i}^{\dagger}\hat{s}_{i}-\hat{p}_{i}^{\dagger}\hat{p}_{i})\bigg]-\sum_{i=2}^{L}t_{ps}\hat{s}_{i}^{\dagger}\hat{p}_{i-1}+\nonumber\\&\sum_{i=1}^{L-1}\bigg[-t(\hat{s}_{i}^{\dagger}\hat{s}_{i+1}-\hat{p}_{i}^{\dagger}\hat{p}_{i+1})+t_{ps}\hat{s}_{i}^{\dagger}\hat{p}_{i+1}\bigg]+\textrm{H.c.}
    \label{eq:Ham2}
\end{align}
\begin{figure}[h]
		\includegraphics[width=0.5\textwidth]{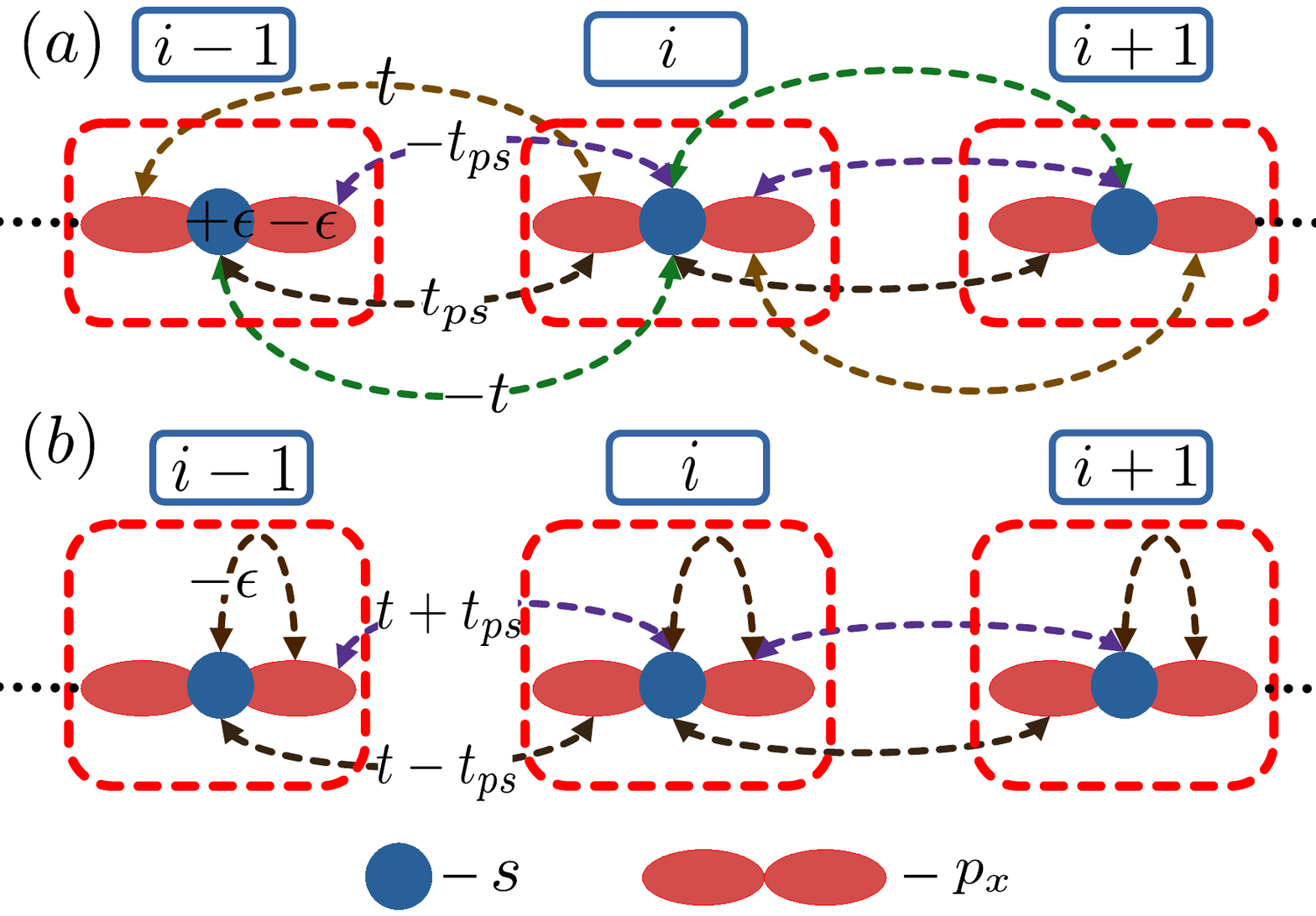}
		\caption{(Color online) Pictorial representations of (a) the model represented by the Hamiltonian $H_1$ (equation \eqref{eq:Ham2}) (b) the model represented by the Bloch Hamiltonian $h'_1(k)$ (equation \eqref{eq:kspace_5}). The orbitals, $s$ and $p_x$, the hopping amplitudes ($t$ and $t_{ps}$) and the onsite energy ($\epsilon$) are shown.} 
    \label{model_1_H}
\end{figure}
after the Fourier transformation of equation \eqref{eq:Ham2}, the Bloch Hamiltonian takes the form,
\begin{equation}
		h_1(k)=\begin{pmatrix}
			\epsilon-2t\cos k & 2it_{ps}\sin k\\-2it_{ps}\sin k & -\epsilon+2t\cos k
		\end{pmatrix}.
		\label{eq:kspace_2}
\end{equation}
This implies a staggered potential of strength $\epsilon$ at each of the orbitals $s$ and $p_x$, hopping amplitudes of $t$ ($s^i\leftrightarrow s^{i+1}$, $p_x^i\leftrightarrow p_x^{i+1}$) and $t_{ps}$ ($s^i\leftrightarrow p_x^{i+1}$), shown in figure \ref{model_1_H}(a).
It is well known that the topological phases depend on the various symmetries of the system, which can be obtained by observing whether the corresponding symmetry operators commute (or anti-commute) with the Bloch Hamiltonian.
In the present case, the system has TRS, which is given by the following relation for any $h(k)$ \cite{PhysRevB.78.195125},
\begin{equation}
    \mathcal{T_+}h(k)\mathcal{T_+}^{-1}=h(-k);\quad\mathrm{with}\quad\mathcal{T_+}\mathcal{T_+^*}=\pm1,
    \label{eq:TRS}
\end{equation}
where $\mathcal{T}$ ($=U_T\mathcal{K}$, $U_T$ being a unitary matrix) is the TRS operator that is anti-unitary in nature, and $h_1(k)$ satisfies equation \eqref{eq:TRS}.
For systems consisting of spinless fermions (present case), $\mathcal{T}$ is nothing but the complex conjugation operator $\mathcal{K}$ and $\mathcal{T}^2=1$.
This implies that if the Hamiltonian $H_1$ ($h_1(k)$) has an eigenvalue $E$ ($E_k$) corresponding to the eigenvector $\ket{\psi}$ ($\ket{\psi(k)}$), then there also exists an eigenvalue $E^*$ ($E_{-k}^*$) corresponding to the eigenvector $\ket{\psi}^*$ ($\ket{\psi(-k)}^*$).
In the same way, the system also possesses particle-hole symmetry (PHS), which is written as,
\begin{equation}
		\mathcal{C}h(k)\mathcal{C}^{-1}=-h(-k);\quad\mathrm{with}\quad\mathcal{C}^2=\pm1.
		\label{eq:PHS}
\end{equation}
The anti-unitary PHS operator $\mathcal{C}=U_C\mathcal{K}$ ($U_C$ is unitary) anti-commutes with the Bloch Hamiltonian $h_1(k)$ with $U_C=\sigma_x$.
Thus, the eigenvalues come in pairs, $\pm E$, with corresponding eigenvectors being $\ket{\psi}$ and $\mathcal{C}\ket{\psi}$.
The presence of Chiral Symmetry (CS) is also noted, as the CS operator ($\Gamma$) is nothing but $\Gamma=\mathcal{T}\cdot\mathcal{C}$.
These characteristics lead us to classify the system within the class $\pmb{\mathrm{BDI}}$ in $\pmb{\mathrm{AZ}}$ symmetry class \cite{PhysRevB.55.1142}.

\begin{figure}[h]
		\includegraphics[width=0.5\textwidth, height=0.6\columnwidth]{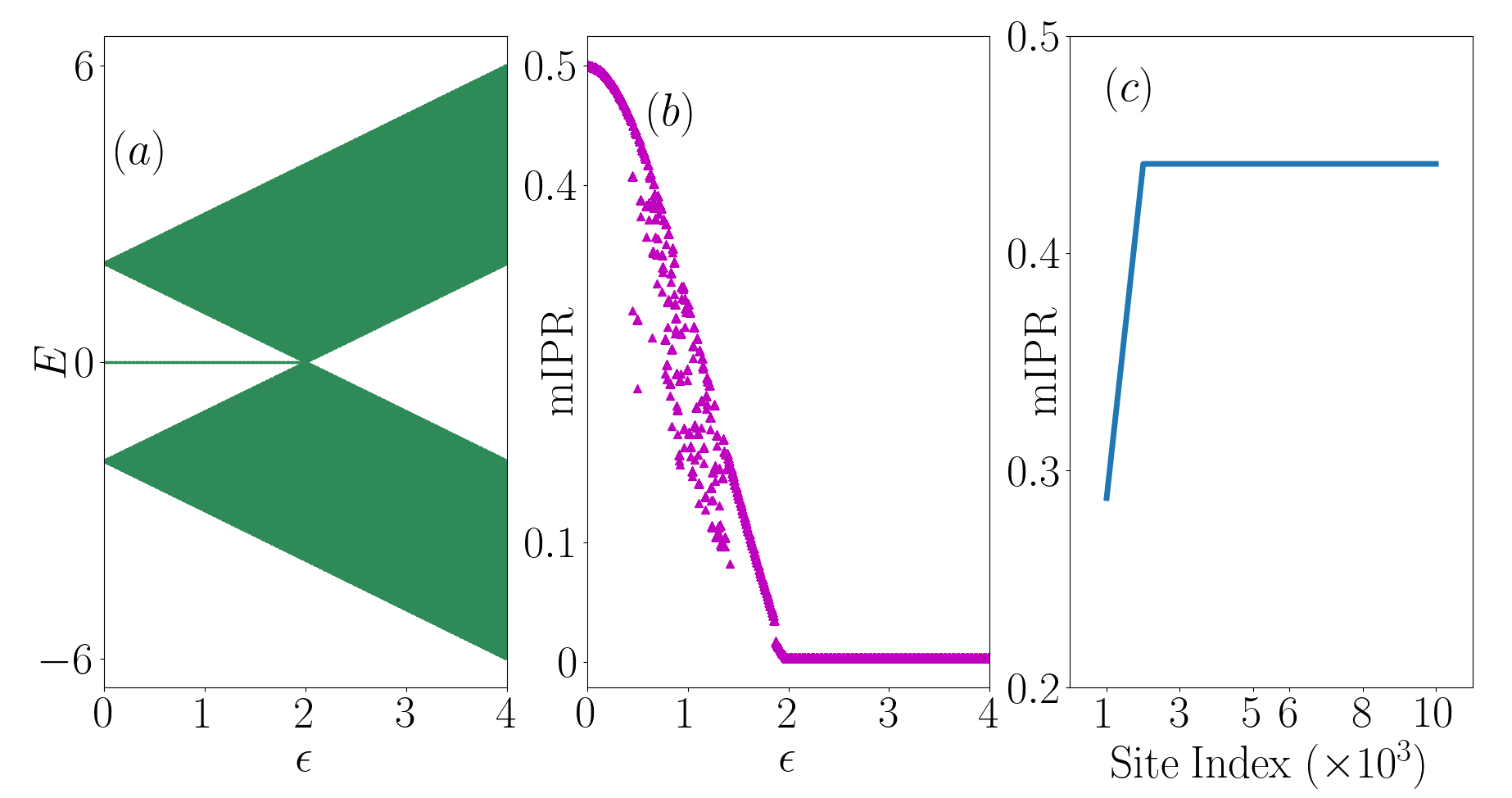}
		\caption{(Color online) (a) Eigenvalues of $H_1$ vs $\epsilon$, showing the appearance(disappearance) of zero energy eigenstates prior to(beyond) $\epsilon=2t$, (b) mIPR vs $\epsilon$ is shown, suggesting that the edge states are non-existent after the point $\epsilon=2t$, and (c) system size effects on mIPR with $\epsilon=0.5t$.}
    \label{rs_1}
\end{figure}

Next, we explore the topological properties and edge state behaviors of this model with OBC and calculate the topological invariants for each model for PBC.
First, we take the Hermitian model represented by equation \eqref{eq:Ham2} for a 1D chain of length $L$.
In figure \ref{rs_1}(a), we have presented the real space eigenspectra with the onsite energy $\epsilon$ being varied from $0$ to $4t$.
Here, we have set $t=t_{ps}=1$ and, in most cases, set the energy scale to a unit of $t$.
The existence of a two-fold degenerate zero energy edge state till $\epsilon=2t$ and their disappearance beyond that point suggest a topological phase transition at that point.
To ascertain the localization of the zero energy edge states, we use the familiar approach of computing the IPR \cite{BKramer_1993}, defined via,
\begin{equation}
		\mathrm{IPR}^{(i)}=\frac{\sum_n|\psi^i_n|^4}{\left(\sum_n|\psi^i_n|^2\right)^2},
    \label{eq:IPR}
\end{equation}
where $\mathrm{IPR}^{(i)}$ is the IPR of the $i$-th eigenstate and $n$ denotes site index.
It is well established that the IPR varies inversely with the system size ($\sim L^{-1}$) for extended states.
In contrast, the IPR becomes independent of the system size for the localized states and approaches $1$ (in the thermodynamic limit) when they are completely localized at any site.
Here, the maximum IPR (mIPR) represents the IPR of the edge states.
Further, it also denotes the highest value of IPR among all the eigenstates for the trivial phase, which does not have edge states.
It is plotted as a function of the potential, $\epsilon$, in figure \ref{rs_1}(b).
The non-zero values suggest that the edge states exist until $\epsilon=2t$ and vanish afterwards.
Figure \ref{rs_1}(c) shows the system size effects on mIPR, suggesting that it is independent of the system size beyond a critical system size of $L\sim1000$.
Thus, we have fixed the number of unit cells at $1000$ for the numerical analysis of mIPR.

The BBC implies that some topological invariant must exist, which will aid in extracting information about the edge (OBC) through the information obtained from the bulk (PBC).
We shall now do the calculations in the momentum space (PBC) and compute the values of the topological invariant.
Such a useful quantity pertaining to particular topological phases of matter is the Berry phase \cite{M_V_Berry_1985}.
It is a geometric phase acquired by the eigenstates over an adiabatic cycle in the parameter space, such as time, position, momentum, etc.
We shall use Berry phase as the suitable invariant throughout the paper, whose definition is given by,
\begin{equation}
		\gamma_{\pm}=i\oint_{BZ}\bra{\psi_{\pm}(k)}\nabla_k\ket{\psi_{\pm}(k)}\;dk,
    \label{berry}
\end{equation}
where $\ket{\psi_+(k)}$ ($\ket{\psi_-(k)}$) is the eigenvector corresponding to the upper (lower) band of the Bloch Hamiltonian, and $\pm$ signs label the band index.
Generally, the Berry phase is an integer (positive or negative) in the unit of $\pi$.

$h_1(k)$ in equation \eqref{eq:kspace_2} can be written as a Dirac Hamiltonian, $h_1(k)=\pmb{d_1\cdot\sigma}$, where $\pmb{d_1}\equiv(0, -2t_{ps}\sin k, \epsilon-2t\cos k)$ and $\pmb{\sigma}$ denote the Pauli matrices.
The presence of the $d_{1z}$ term in the $\pmb{d_1}$-vector poses a challenge for computing the Berry phase from equation \eqref{berry}.
Hence, to make $d_{1z}$ zero, we shall perform a unitary transformation on $h_1(k)$.
This is achieved via a unitary matrix $U$, such that,
\begin{gather}
		h_1'(k)=U^{\dagger}h_1(k)U;\quad \mathrm{where}\\ U=\frac{1}{\sqrt{2}}
		\begin{pmatrix}
			1 & -1\\1 & 1
		\end{pmatrix},
		\label{eq:U}
	\end{gather}
	which yields,
	\begin{align}
		h_1'(k)=
		\begin{pmatrix}
			0 & -\epsilon+2t\cos k\\ &+2it_{ps}\sin k\\-\epsilon+2t\cos k\\-2it_{ps}\sin k & 0
		\end{pmatrix},
    \label{eq:kspace_5}
\end{align}
\begin{figure}[t]
		\includegraphics[width=0.5\textwidth, height=0.55\columnwidth]{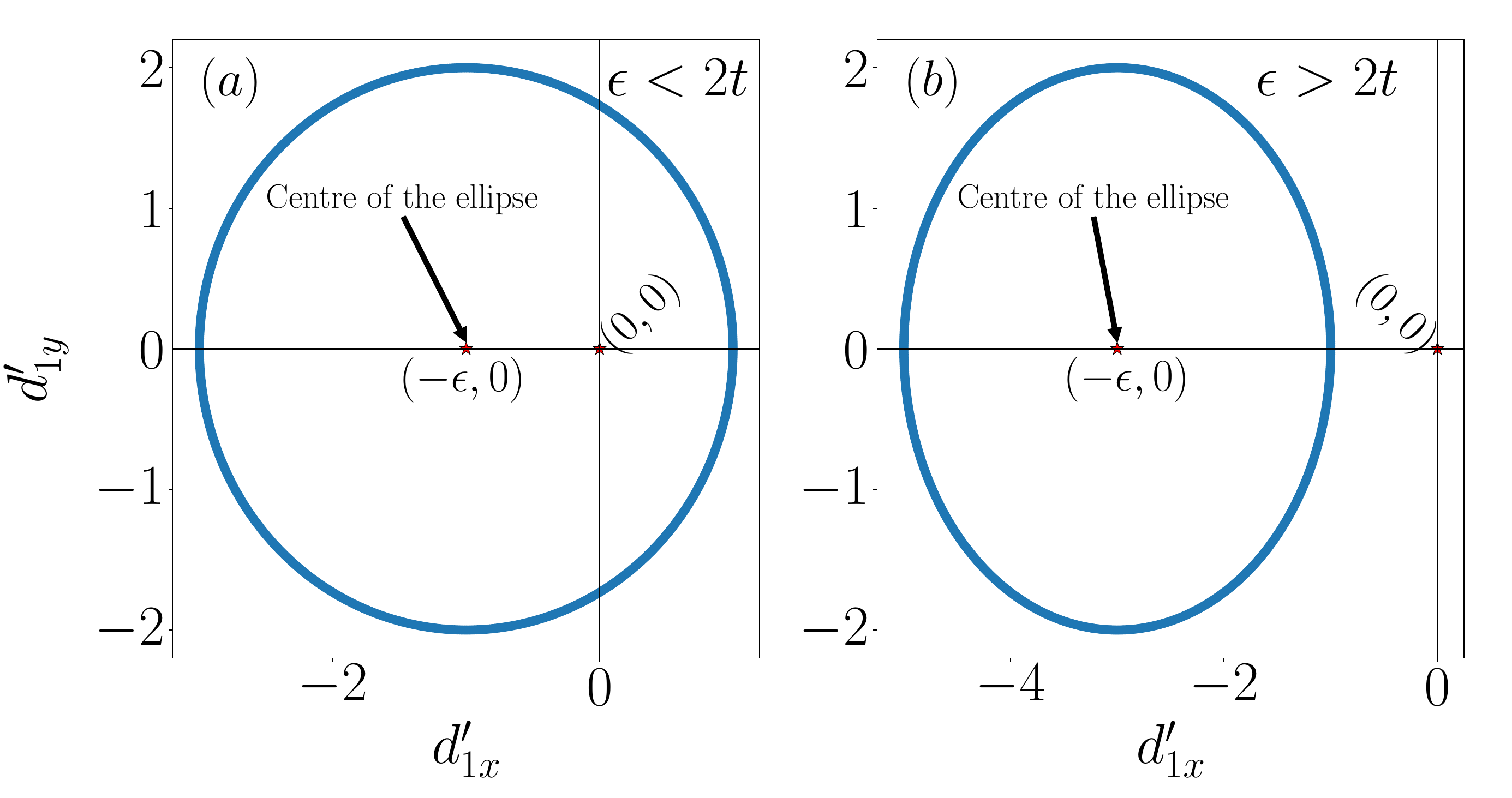}
		\caption{(Color online) The locus of $\pmb{d'_1}$-vector (an ellipse) in $d'_{1x}-d'_{1y}$ plane for the two cases, namely, (a) $\epsilon<2t$, where the origin $(0,0)$ is inside the closed loop, and (b) $\epsilon>2t$, where the origin is outside the ellipse. The center of the ellipse is at $(-\epsilon,0)$.}
    \label{dxdy_1}
\end{figure}
The transformed Bloch Hamiltonian $h'_1(k)$ corresponds to a different two-orbital model shown in figure \ref{model_1_H}(b), which has $-\epsilon$ and $t+t_{ps}$($t-t_{ps}$) as the hopping amplitudes for $s^i\leftrightarrow p_x^i$ and $p_x^i\leftrightarrow s^{i+1}$($s^i\leftrightarrow p_x^{i+1}$) respectively.
However, it falls under the same symmetry class, namely $\pmb{\mathrm{BDI}}$, which ensures that it has all the symmetries which the original model (represented by $h_1(k)$) possesses.
This is obvious because they are connected via a similarity transformation, and hence, the band structure remains identical.
Hence, they must be `topologically equivalent'.

The Dirac form for $h'_1(k)$ is given as, $h'_1(k)=\pmb{d'_1\cdot\sigma}$ with $\pmb{d'_1}\equiv(-\epsilon+2t\cos k, -2t_{ps}\sin k, 0)$.
The unitary transformation interchanges the $x$ and the $z$ components of the $\pmb{d_1}$-vector in addition to rendering a negative sign, that is, $d'_{1x}=-d_{1z}$ and $d'_{1z}=-d_{1x}$.
In the $d'_{1x}-d'_{1y}$ plane, the $\pmb{d'_1}$-vector forms a loop (see figure \ref{dxdy_1}) as $k$ goes from $-\pi$ to $+\pi$ in the Brillouin zone (BZ) and includes the origin ($0,0$), about which the trajectory of the $\pmb{d'_1}$-vector can be seen to look for the topological properties.
For the case $\epsilon<2t$, the EP is enclosed (figure \ref{dxdy_1}(a)) and excluded (figure \ref{dxdy_1}(b)) when $\epsilon>2t$.

\begin{figure}[t]
    \includegraphics[width=0.5\textwidth, height=0.55\columnwidth]{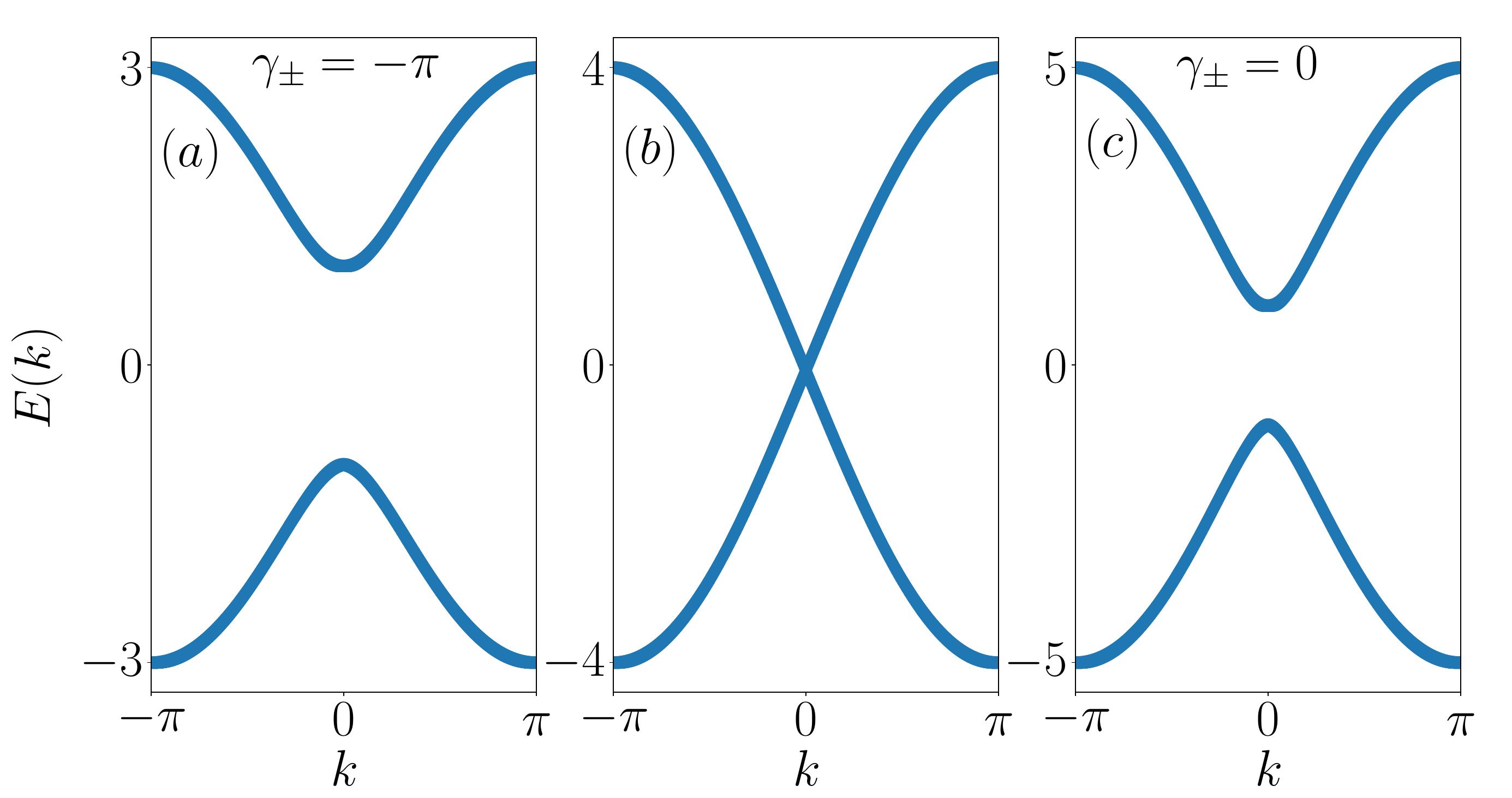}
		\caption{(Color online) The energy dispersion corresponding to $H_1$ ($h_1(k)$ in $k$-space) shown for the three cases, namely, (a) $\epsilon<2t$ (gapped), (b) $\epsilon=2t$ (where the spectral gap closes), and (c) $\epsilon>2t$ (gap reopens) keeping $t=t_{ps}=1$. (a) and (c) imply the topological and the trivial phases of $H_1$, whereas (b) shows a phase transition point.}
    \label{band_1}
\end{figure}

The expression for the energy, $E_1(k)$ corresponding to Bloch Hamiltonians $h_1(k)$ or $h'_1(k)$, is given by,
\begin{equation}
		E_{1\pm}(k)=\pm\sqrt{(-\epsilon+2t\cos k)^2+4t_{ps}^2\sin^2k},
    \label{bs_1}
\end{equation}
and is plotted in figure \ref{band_1}.
The first column (figure \ref{band_1}(a)) and the third column (figure \ref{band_1}(c)) show gapped eigenspectra corresponding to the cases $\epsilon<2t$ and $\epsilon>2t$, respectively, but with different values of Berry phase which we will discuss later in this section.
They correspond to the scenario depicted in figure \ref{dxdy_1}(a) and figure \ref{dxdy_1}(b), respectively.
The spectral gap vanishes at $k=0$ in figure \ref{band_1}(b), which corresponds to $\epsilon=2t$, which is the case when the EP ($0,0$) lies on the locus of the $\pmb{d'_1}$-vector.
\begin{figure}[b]
    \includegraphics[width=0.5\textwidth, height=0.6\columnwidth]{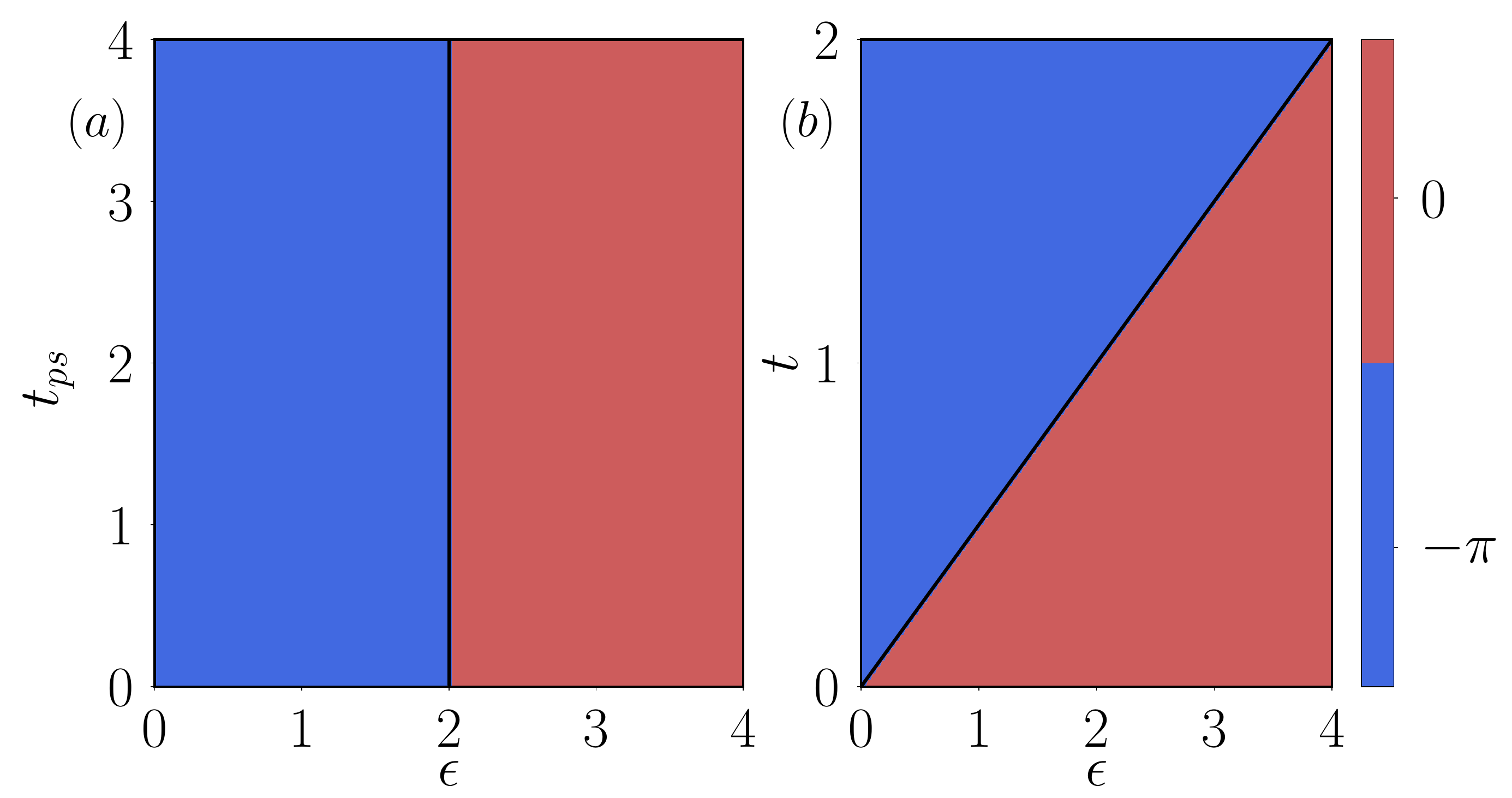}
		\caption{(Color online) The phase diagram obtained via computing the Berry phase for the parameters (a) $\epsilon$ and $t_{ps}$ with $t=1$, and (b) $\epsilon$ and $t$, with $t_{ps}=1$. The phase diagrams show that the topological phase transition depends only on $\epsilon$ and $t$ regardless of the value of $t_{ps}$, and it occurs at $\epsilon=2t$. In the region, where $\epsilon<2t$, the Berry phase acquires a value $-\pi$ and $0$ for $\epsilon>2t$.}
    \label{B_phase}
\end{figure}

To calculate the Berry phase, we shall use the eigenvectors of $h'_1(k)$ corresponding to the eigenvalues $E_{1\pm}(k)$, which are given by,
\begin{equation}
		\ket{\psi'_{1\pm}(k)}=\pm\frac{1}{\sqrt{2}}e^{\pm i\zeta}
		\begin{pmatrix}
			e^{-i\phi_{1k}}\\1
    \end{pmatrix},
\end{equation}
where $\zeta$ is independent of $k$ and $\phi_{1k}$ is given by,
\begin{equation}
		\phi_{1k}=\tan^{-1} \left(\frac{-2t_{ps}\sin k}{-\epsilon+2t\cos k}\right).
\end{equation}
Putting $\ket{\psi'_{1\pm}(k)}$ in equation \eqref{berry}, we will get the expression of the Berry phase, $\gamma_{\pm}$, given by,
\begin{align}
		\gamma_{\pm}&=\frac{1}{2}\oint_{BZ}\frac{\partial\phi_{1k}}{\partial k}\;dk\nonumber\\&=\frac{1}{2}\oint_{BZ}\frac{2t_{ps}(\epsilon\cos k-2t)}{(-\epsilon+2t\cos k)^2+4t_{ps}^2\sin^2k}\;dk.
		\label{eq:berry1}
\end{align}

To obtain a phase diagram for the Berry phase using equation \eqref{eq:berry1}, we have considered the parameter space spanned by the onsite potential $\epsilon$ and the hopping strength $t_{ps}$, keeping $t$ fixed at $1$.
As can be seen in figure \ref{B_phase}(a), there is a sharp transition in the value of the Berry phase depending on the values of $\epsilon$ and $t$, however, it is independent of $t_{ps}$.
Similarly, figure \ref{B_phase}(b) shows that the transition occurs along a straight line whose equation is $\epsilon=2t$.
Above the line ($\epsilon<2t$), the Berry phase acquires a value $-\pi$ and below ($\epsilon>2t$) it vanishes.
These are reminiscent of the appearance and disappearance of two zero energy edge states (shown in figure \ref{rs_1}(a)) and can be identified as the topological and trivial phases, respectively.
	
\section{\label{sec3}Non-$\mathcal{PT}$ symmetric NH models}

Next, we explore the NH extensions of the Hermitian model described by equation \eqref{eq:Ham2}.
The well-adapted (and usually employed) ways to convert a Hermitian model to an NH one can be done via,
\begin{enumerate}[label=\roman*., itemsep=0pt, topsep=0pt]
		\item introducing imaginary onsite potential (or (complex) random disorder).
		\item breaking the reciprocity in the hopping strengths between the lattice sites.
\end{enumerate}
We will explore models that lack the $\mathcal{PT}$ symmetry in the subsequent subsections.
Furthermore, we have categorized non-$\mathcal{PT}$ symmetric NH models into two groups, namely, those with and without TRS, and examined their topological and localization properties in details.

\subsection{Non-$\mathcal{PT}$ symmetric NH model without TRS}

In this section, we only introduce a staggered imaginary onsite potential in the system.
This will result in a Hamiltonian of the form given by,
\begin{align}
		H_2&=\sum_{i=1}^{L}\bigg[i\epsilon(\hat{s}_{i}^{\dagger}\hat{s}_{i}
		-\hat{p}_{i}^{\dagger}\hat{p}_{i})\bigg]-\sum_{i=2}^{L}t_{ps}\hat{s}_{i}^{\dagger}\hat{p}_{i-1}+\nonumber\\&\sum_{i=1}^{L-1}\bigg[-t(\hat{s}_{i}^{\dagger}\hat{s}_{i+1}-\hat{p}_{i}^{\dagger}\hat{p}_{i+1})+t_{ps}\hat{s}_{i}^{\dagger}\hat{p}_{i+1}\bigg]+\textrm{H.c.},
    \label{eq:Ham3}
\end{align}
where $\epsilon$ is real and denotes the magnitude of the imaginary onsite potential.
The corresponding Bloch Hamiltonian is given by,
\begin{equation}
    h_2(k)=\begin{pmatrix}
			i\epsilon-2t\cos k & 2it_{ps}\sin k\\-2it_{ps}\sin k & -i\epsilon+2t\cos k
		\end{pmatrix}.
    \label{eq:kspace_3}
\end{equation}
The $i\epsilon$ term will break TRS and hence will not possess $\mathcal{PT}$ symmetry, which can be realized via,
\begin{equation*}
		(\mathcal{PT})h_2(k)(\mathcal{PT})^{-1}=-h_2^*(k)\neq h_2(k),
\end{equation*}
where the $\mathcal{PT}$ operator is equal to $\sigma_x\mathcal{K}$ ($\sigma_x:x$ component of Pauli matrices and $\mathcal{K}:$ complex conjugation operator).
It is known that in NH systems, there exists an $\pmb{\mathrm{AZ}^{\dagger}}$ symmetry class \cite{Ashida2020}, apart from the $\pmb{\mathrm{AZ}}$ symmetry class, since $H^*\neq H^T$.
Thus, this system possesses another symmetry $\mathrm{TRS}^{\dagger}$ from the $\pmb{\mathrm{AZ}^{\dagger}}$ class, which demands,
\begin{equation*}
	\mathcal{C}_+h^T(k)\mathcal{C}_+^{-1}=h(-k);\quad\mathrm{with}\quad\mathcal{C}_+\mathcal{C}_+^*=\pm1,
\end{equation*}
for any $h(k)$, with $\mathcal{C}_+$ being an unitary matrix.
Further, $h_2(k)$ satisfies $h_2^T(k)=h_2(-k)$.
So, the present model falls in the class $\pmb{\mathrm{AI}^{\dagger}}$ in the real $\pmb{\mathrm{AZ}^{\dagger}}$ symmetry classification due to the $\mathrm{TRS}^{\dagger}$ symmetry present in the system.

We now analyze the topological and localization properties of the edge modes pertaining to the NH model given by the equations \eqref{eq:Ham3} and \eqref{eq:kspace_3}.
The potential $\pm i\epsilon$ on the $s$ and $p_x$ orbitals in the Hamiltonian $H_2$ physically imply `gain' and `loss' of energy that the system experiences due to the non-Hermiticity.
\begin{figure}[h]
    \includegraphics[width=0.5\textwidth, height=0.6\columnwidth]{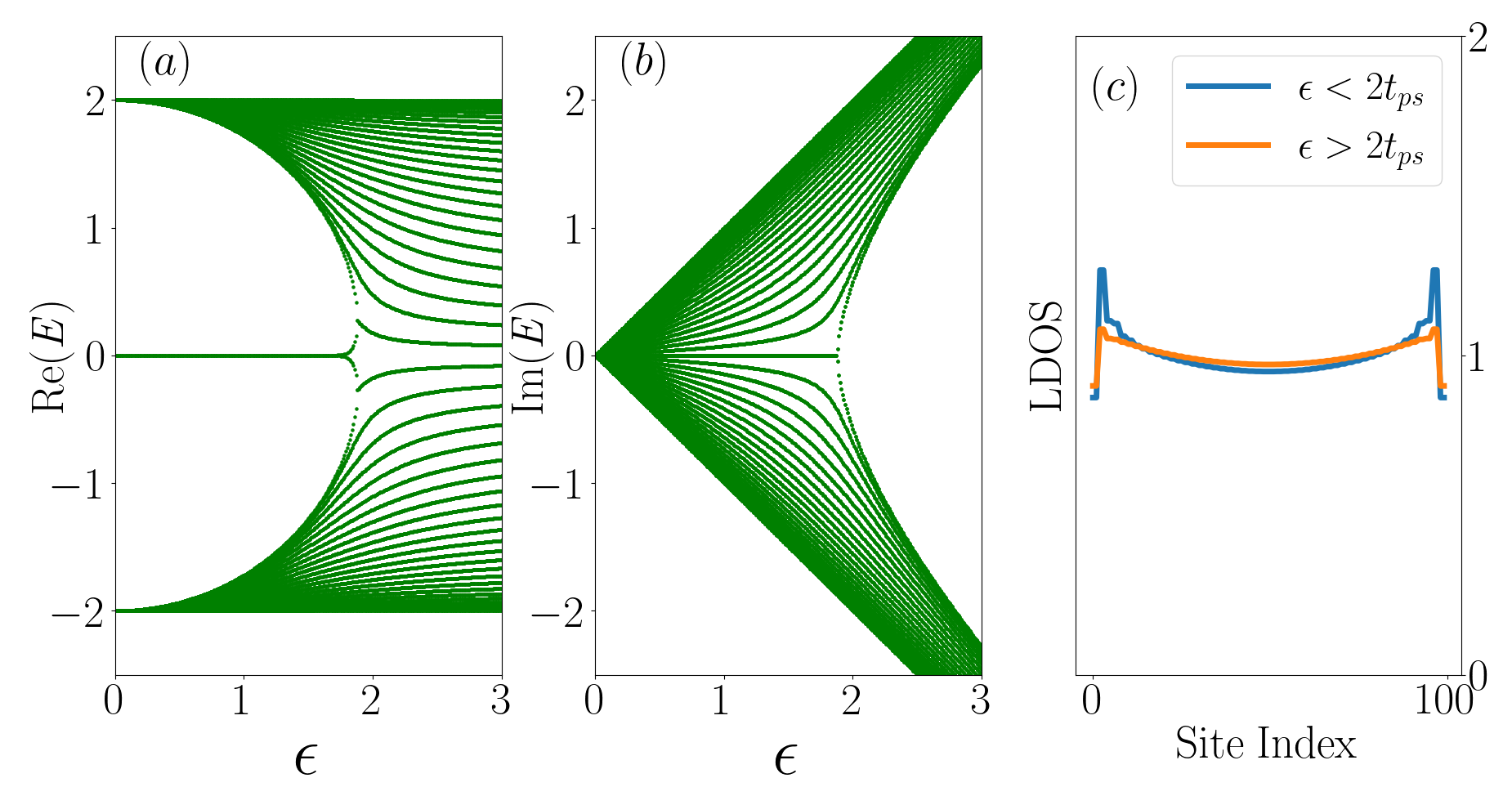}
		\caption{(Color online) The (a) real part and (b) imaginary part of the eigenvalues of $H_2$ are plotted as a function of $\epsilon$ for $50$ unit cells. (b) shows the absence of purely real eigenvalues and hence is suggestive of a non-$\mathcal{PT}$ symmetric case. (a) and (b) show the appearance (disappearance) of doubly degenerate absolute zero energy edge modes, that is, both the real and imaginary parts of the eigenvalues being zero, for $\epsilon<2t_{ps}$ ($\epsilon>2t_{ps}$). (c) LDOS is calculated for both the topological and the trivial phases. The plot confirms the absence of NHSE.}
    \label{rs_2}
\end{figure}
Figure \ref{rs_2}(a), representing the variation of the real part (Re($E$)) of the eigenvalues, $E$, with the onsite potential, $\epsilon$, suggests that doubly degenerate zero energy edge modes that exist till $\epsilon=2$ (in units of $t_{ps}$), and disappear beyond that.
The presence of non-zero imaginary parts in the energy spectrum depicted in Figure \ref{rs_2}(b) serves as an evidence for the absence of $\mathcal{PT}$ symmetry of the system.
Figures \ref{rs_2}(a) and \ref{rs_2}(b) demonstrate that the energy eigenvalues, $E$, appear in pairs, that is $\pm E$, which is a manifestation of PHS.
In figure \ref{rs_2}(c), we have shown the local density of states (LDOS), which can be obtained from the expression, $$\mathrm{LDOS}^{(n)}=\sum_{i=1}^{2L}|\psi_n^i|^2$$ corresponding to the $n^{\mathrm{th}}$ lattice site, where the summation index, $i$, runs over all eigenenergies.
Higher values of LDOS at the edges of the system for the case $\epsilon<2t_{ps}$ supports the existence of the two localized edge modes, which contrasts the case corresponding to $\epsilon>2t_{ps}$.
It is clear from figure \ref{rs_2}(c) that the NHSE is absent in this system as there is no build-up of LDOS at any of the edges.
\begin{figure}[h]
    \includegraphics[width=0.5\textwidth, height=0.65\columnwidth]{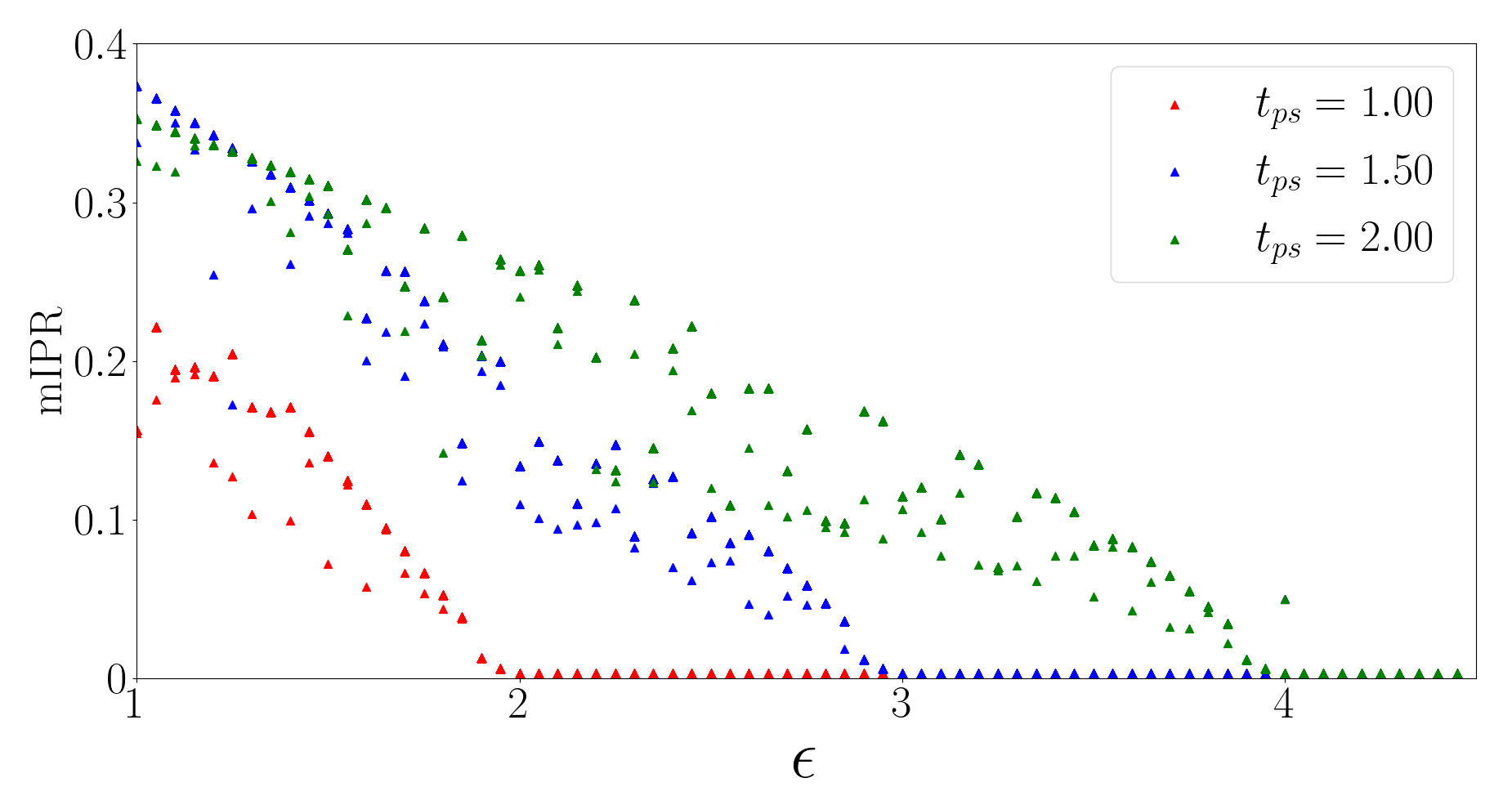}
		\caption{(Color online) The maximum value of IPR, mIPR (defined in the text), versus $\epsilon$ is plotted here for different values of $t_{ps}$ keeping $t=1$. The plot suggests that the existence of the edge states now depends on $t_{ps}$. The scenario is unlike the Hermitian model, where it depends on $t$.}
    \label{ipr1}
\end{figure}

For a generic NH Hamiltonian (say $H_{\mathrm{NH}}$), the right eigenvector, that is, the eigenvectors of $H_{\mathrm{NH}}$ and left eigenvector (the eigenvectors of $H_{\mathrm{NH}}^{\dagger}$) corresponding to a particular eigenvalue differ from each other.
The orthonormality condition in this case is given by, $$\left<\lambda_m|\psi_n\right>=\delta_{mn},$$ where $\ket{\lambda_m}$ and $\ket{\psi_n}$ are the left and the right eigenvectors corresponding to $m^{th}$ and $n^{th}$ eigenvalues of $H_{\mathrm{NH}}$ \cite{PhysRevLett.121.026808, PhysRevB.97.045106}.
The IPR is given by the same formula as in equation \eqref{eq:IPR}, except that the right eigenvectors will have to be used for this case.
The mIPR, defined in the previous section, is computed as a function of $\epsilon$ for three different values of $t_{ps}$, namely, $t_{ps}=1,1.5,2.0$ and is shown in figure \ref{ipr1}.
The presence of the zero energy edge states for the model depends on $t_{ps}$.
The mIPR is non-zero (edge states) for values of $\epsilon$ lower than $2t_{ps}$, and beyond that, it is zero, implying that the edge states exist till $\epsilon=2t_{ps}$, and disappear beyond that.

Next, we analyze the scenario with PBC.
We perform the same unitary transformation as done in the previous section, given in equation \eqref{eq:U}, on the Bloch Hamiltonian, $h_2(k)$ in equation \eqref{eq:kspace_3} which yields,
\begin{align}
		h_2'(k)=
		\begin{pmatrix}
			0 & -i\epsilon+2t\cos k\\ &+2it_{ps}\sin k\\-i\epsilon+2t\cos k\\-2it_{ps}\sin k & 0
		\end{pmatrix},
    \label{eq:kspace_6}
\end{align}
which can be written in a Dirac form as, $h'_2(k)=\pmb{d'_2}\cdot\pmb{\sigma}$ with $\pmb{d'_2}\equiv(-i\epsilon+2t\cos k, -2t_{ps}\sin k, 0)$.
Evidently, $\pmb{d'_2}$ has all the symmetries, that is, both PHS and $\textrm{TRS}^{\dagger}$, which eventually says that $\pmb{d_2}$ and $\pmb{d'_2}$ have the same eigenspectra, and are given by,
\begin{equation}
		E_{2\pm}(k)=\pm\sqrt{(-i\epsilon+2t\cos k)^2+4t_{ps}^2\sin^2k}.
    \label{bs_2}
\end{equation}
The left ($\ket{\lambda'_{2\pm}(k)}$) and the right ($\ket{\psi'_{2\pm}(k)}$) eigenfunctions of $h'_2(k)$ are given by,
\begin{align}
		\ket{\lambda'_{2\pm}(k)}&=\frac{1}{\sqrt{2}}e^{\pm i\eta^*}
		\begin{pmatrix}
			\pm e^{-i\phi_{2k}^*}\\1
		\end{pmatrix};\nonumber\\
		\ket{\psi'_{2\pm}(k)}&=\frac{1}{\sqrt{2}}e^{\pm i\eta}
		\begin{pmatrix}
			\pm e^{-i\phi_{2k}}\\1
		\end{pmatrix}.
    \label{eig_v_2}
\end{align}
Here $\eta$ is a constant (and independent of $k$ and $\phi_{2k}$), and $\phi_{2k}$ is given by,
\begin{equation*}
		\phi_{2k}=\tan^{-1} \left(\frac{-2t_{ps}\sin k}{-i\epsilon+2t\cos k}\right).
\end{equation*}
\begin{figure}[ht]
    \includegraphics[width=0.5\textwidth, height=0.7\columnwidth]{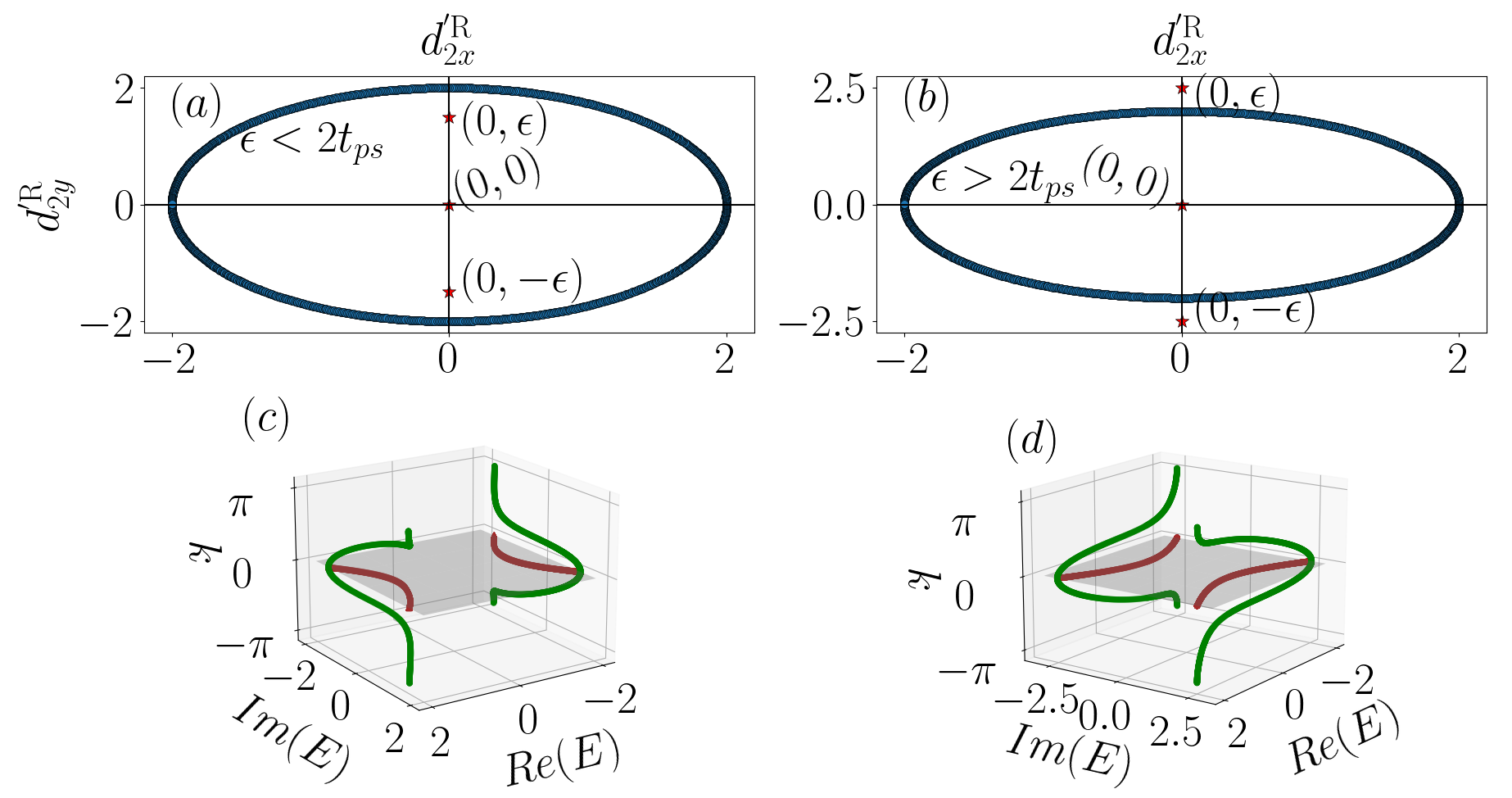}
		\caption{(Color online) Real part of $\pmb{d'_2}$-vector is drawn in a plane spanned by $d_{2x}^{'\mathrm{R}}-d_{2y}^{'\mathrm{R}}$ for the parameters $t=1$, $t_{ps}=1$, and (a) $\epsilon=1.5$, where the ellipse encloses the EPs (topological case), and (b) $\epsilon=2.5$, where it excludes the EPs (trivial case). (c) and (d) represent 3D figures of the band structures ($E_{2\pm}(k)$ vs $k$) with the real and the imaginary parts of the energy being plotted along the $x$ and $y$ axis, respectively. At the same time, the momentum $k$ is along the $z$ direction. The points in green and brown denote the actual data points and their 2D projections, respectively.}
    \label{band2}
\end{figure}
\begin{figure}[hb]
    \includegraphics[width=0.5\textwidth, height=0.6\columnwidth]{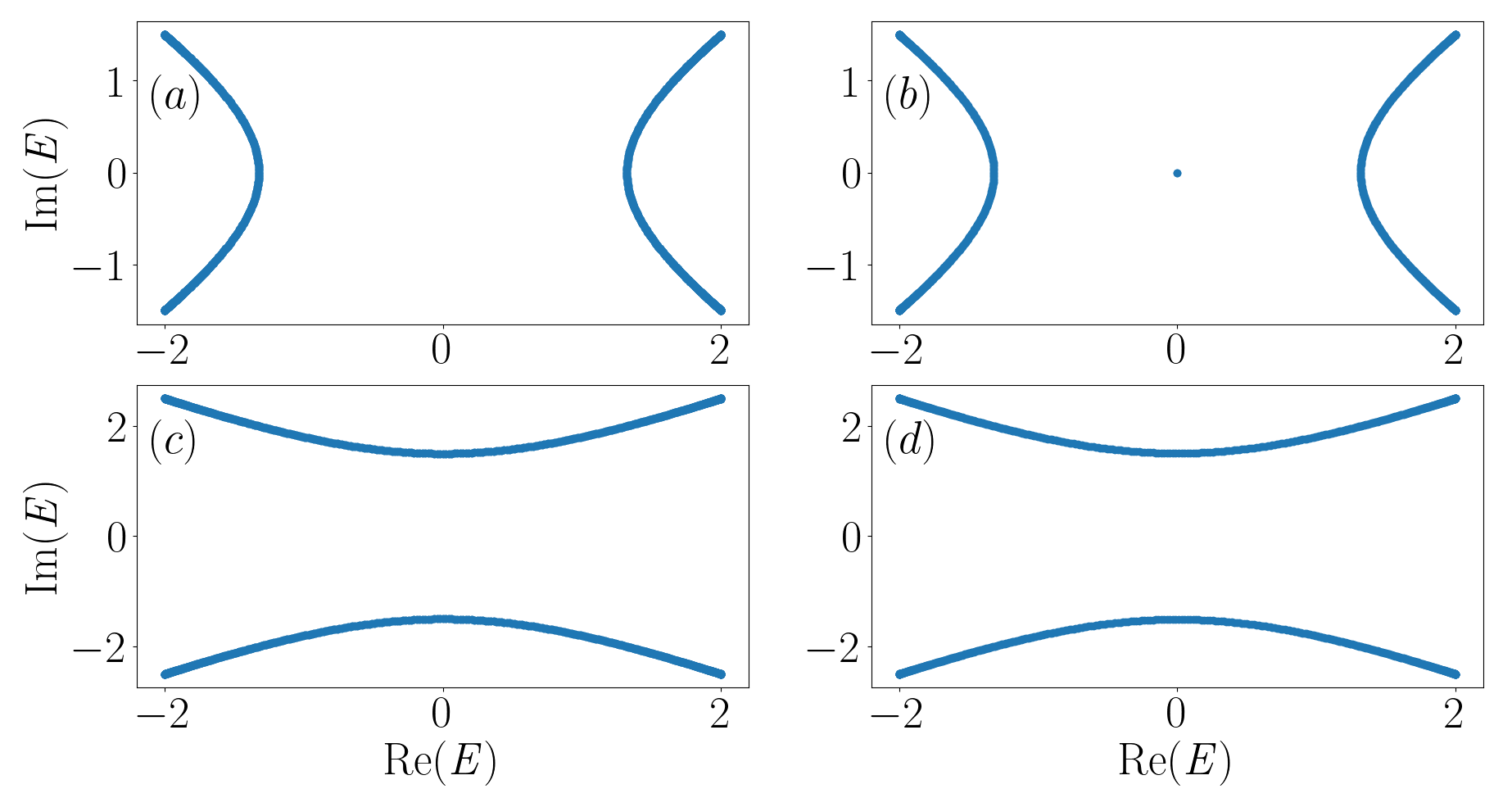}
		\caption{(Color online) The real and the imaginary parts of energy, corresponding to the Hamiltonian $H_2$, for PBC (left panel) and OBC (right panel) are shown. The parameters are kept as follows: $\epsilon<2t_{ps}$ for both (a) and (b), while $\epsilon>2t_{ps}$ for both (c) and (d).}
    \label{rm_i}
\end{figure}

Due to the non-Hermiticity of $h'_2(k)$, $\pmb{d'_2}$ is also complex unlike $\pmb{d'_1}$.
With the help of a bit of algebra, one can show that the locations of the two EPs in a space spanned by $d_{2x}^{'\mathrm{R}}-d_{2y}^{'\mathrm{R}}$ are at ($d_{2y}^{'\mathrm{I}}, -d_{2x}^{'\mathrm{I}}$) and ($-d_{2y}^{'\mathrm{I}}, d_{2x}^{'\mathrm{I}}$) \cite{Halder_2023, PhysRevA.97.052115} which are ($0, \epsilon$) and ($0, -\epsilon$) respectively, where $d_{2x}^{'\mathrm{R}}$ ($d_{2y}^{'\mathrm{R}}$) and $d_{2x}^{'\mathrm{I}}$ ($d_{2y}^{'\mathrm{I}}$) represent the real and the imaginary parts of $d'_{2x}$ ($d'_{2y}$) respectively.
At these EPs, for $k=\frac{\pi}{2}$, $E_{2}(k)$ vanish and $\left<\lambda'_{2\pm}(k)|\psi'_{2\pm}(k)\right>$ becomes ill-defined.
The locus of the real part of $\pmb{d'_2}$ constitutes an ellipse with the center at ($0,0$) in the $d_{2x}^{'\mathrm{R}}-d_{2y}^{'\mathrm{R}}$ plane as shown in figures \ref{band2}(a) and \ref{band2}(b) corresponding to two representative values of $\epsilon$, namely, $\epsilon=1.5$ and $\epsilon=2.5$ respectively.
The region, where the EPs are inside (outside) the ellipse in figure \ref{band2}(a) (figure \ref{band2}(b)), which denotes $\epsilon<2t_{ps}$ ($\epsilon>2t_{ps}$), and is referred to as the topological (trivial) phase with two (no) zero energy edge modes.
Further, figures \ref{band2}(c) and \ref{band2}(d) are the band structures for the same parameters as used in figures \ref{band2}(a) and \ref{band2}(b), respectively.
The plot shown in green color denotes $E_{2\pm}(k)$ as a function of $k$, and the one in brown color represents the 2D projection of the same on the Re($E$)-Im($E$) plane.
Figure \ref{band2}(c), which corresponds to the topological phase, shows that there is a real line gap in the band structure as $$\mathrm{Re}(E_2(k))\neq 0\;\forall\;k.$$
The base axis is the imaginary axis, with the projected $E_2(k)$ lying on either side of the base axis.
At $\epsilon=2t_{ps}$, the two bands are intertwined at the points, $-\frac{\pi}{2}$, $0$ and $\frac{\pi}{2}$, in the BZ.
Thus, the Hamiltonian becomes gapless at these points, where the topological to the trivial phase transition occurs.
In contrast, in figure \ref{band2}(d), for $\epsilon>2t_{ps}$, the line gap becomes imaginary, with the base axis now being replaced by the real axis, and the system is in a trivial phase.

We also investigate the comparison between PBC and OBC on the eigenspectra.
NHSE is characterized by strong sensitivity of the eigenvalue spectra to the boundary conditions.
The absence of NHSE is provided by the Re($E$) vs Im($E$) plot, presented in figure \ref{rm_i}, for both the PBC (left panel) and the OBC (right panel) in both the topological (upper row) and the trivial (lower row) regions, respectively.
This observation confirms that the eigenvalue spectra are unaffected by the choice of the boundary conditions, except for the presence of two edge modes, present in the system.

The origin of the complex Berry phase (CBP) \cite{GARRISON1988177}, which denotes the topological invariant for the present case, remains equivalent to that for the Hermitian case, with the only difference that now both the left and right eigenvectors have to be employed for computation.
Thus, the definition of the CBP becomes,
\begin{equation}
    \gamma_{\pm}=i\oint_{BZ}\bra{\lambda'_{2\pm}(k)}\nabla_k\ket{\psi'_{2\pm}(k)}\;dk.
    \label{Berry_NH}
\end{equation}
Substituting the expressions of $\ket{\lambda'_{2\pm}(k)}$ and $\ket{\psi'_{2\pm}(k)}$ (equation \eqref{eig_v_2}) in equation \eqref{Berry_NH}, we get the expression for CBP as,
\begin{align}
		\gamma_{\pm}&=\frac{1}{2}\oint_{BZ}\frac{\partial\phi_{2k}}{\partial k}\;dk\nonumber\\&=\frac{1}{2}\oint_{BZ}\frac{2t_{ps}(i\epsilon\cos k-2t)}{(-i\epsilon+2t\cos k)^2+4t_{ps}^2\sin^2k}\;dk
    \label{eq:berry2}
\end{align}
\begin{figure}[h]
    \includegraphics[width=0.5\textwidth, height=0.6\columnwidth]{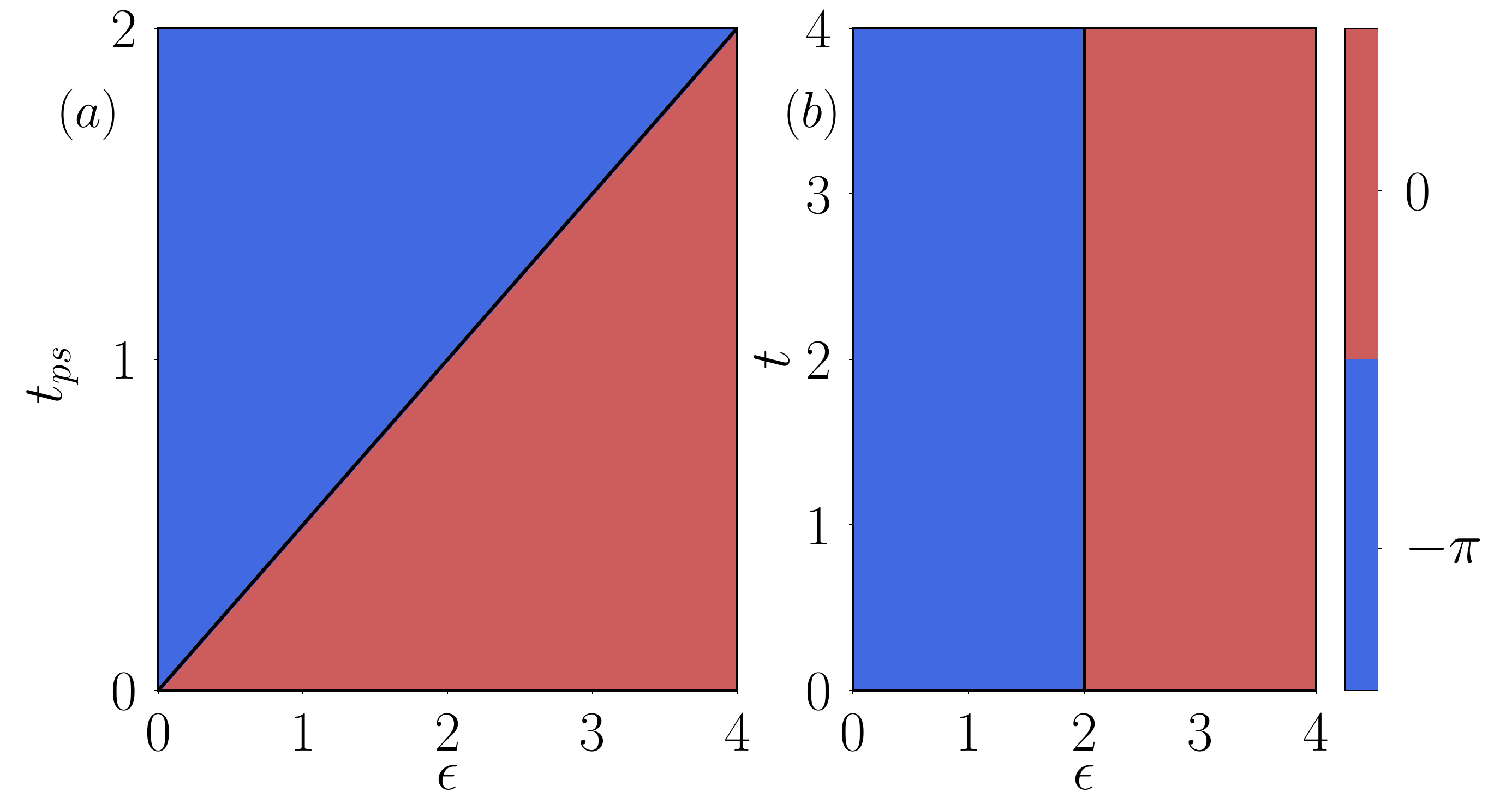}
		\caption{(Color online) Phase diagram based on the value of CBP for the parameters (a) $\epsilon$ and $t_{ps}$ keeping $t=1$ and (b) $\epsilon$ and $t$ keeping $t_{ps}$ fixed at $1$. The phase diagrams suggest that the values of CBP depend only on $\epsilon$ and $t_{ps}$ and are independent of the values of $t$. The CBP takes the values $-\pi$ and $0$ for the regions $\epsilon<2t_{ps}$ and $\epsilon>2t_{ps}$, respectively.}
    \label{B_phase_i}
\end{figure}
From the above equation, we get the phase diagram shown in figure \ref{B_phase_i}, where CBP is computed for a range of $\epsilon$, namely, [$0:4$].
Figure \ref{B_phase_i}(a) suggests that a phase transition occurs across the straight line denoted by $\epsilon=2t_{ps}$, while figure \ref{B_phase_i}(b) depicts the values of CBP in the $\epsilon$-$t$ plane ($t_{ps}=1$).
The plot suggests that CBP is independent of $t$. However, it changes value from $-\pi$ to $0$ at $\epsilon=2t_{ps}$, denoting that these are the topological and the trivial phases, respectively.
These phenomena are almost the same as the Hermitian case (see figure \ref{B_phase}), except for the fact that the roles of $t$ and $t_{ps}$ are interchanged with regard to the topological phase transitions.
Also, the BBBC is preserved here, and NHSE is absent which is somewhat expected.

\subsection{Non-$\mathcal{PT}$ symmetric NH model with TRS}

We proceed to construct another type of NH model that lacks Parity-Time ($\mathcal{PT}$) symmetry but retains TRS.
To accomplish this, we fabricate a model where the non-reciprocity term is included in the hopping amplitudes ($t$) between identical orbitals of adjacent unit cells, that is $s^i\leftrightarrow s^{i+1}$ and $p_x^i\leftrightarrow p_x^{i+1}$ hopping.
The Hamiltonian in real space is given by,
\begin{align}
    H_3=H_1-\sum_{i=1}^{L-1}\left[\Delta(\hat{s}_{i}^{\dagger}\hat{s}_{i+1}-\hat{s}_{i+1}^{\dagger}\hat{s}_{i})+\Tilde{\Delta}(\hat{p}_{i}^{\dagger}\hat{p}_{i+1}-\hat{p}_{i+1}^{\dagger}\hat{p}_{i})\right],
    \label{eq:Ham5}
\end{align}
where $\Delta$ ($\Tilde{\Delta}$) is the non-reciprocity in the hopping between $s$ ($p_x$) orbitals.
The corresponding Bloch Hamiltonian can be written as,
\begin{equation}
    h_3(k)=\begin{pmatrix}
		\epsilon-2t\cos k+2i\Delta\sin{k} & 2it_{ps}\sin k\\-2it_{ps}\sin k & -\epsilon+2t\cos k+2i\Tilde{\Delta}\sin{k}
    \end{pmatrix}.
    \label{eq:kspace_8}
\end{equation}
Now, we discuss two special cases of the equation \eqref{eq:kspace_8}, namely, $(i)\Delta=\Tilde{\Delta}$ and $(ii)\Delta=-\Tilde{\Delta}$.
It becomes evident that the Hamiltonian $h_3(k)$ possesses TRS regardless of the relationship between $\Delta$ and $\Tilde{\Delta}$, yet it lacks inversion (parity) symmetry, thereby, classifying it as non-$\mathcal{PT}$ symmetric.
As a result, the eigenvalues of the Hamiltonian become complex, but they appear in complex conjugate pairs due to its retention of TRS.
Additionally, $h_3(k)$ holds the following relations,
\begin{align*}
    \mathcal{C_-}h_3^T(k,\Delta=\Tilde{\Delta})\mathcal{C_-}&=-h_3(-k,\Delta=\Tilde{\Delta});\quad\mathcal{C_-}\mathcal{C_-^*}=\pm1,\nonumber\\
    \mathcal{T_-}h_3^*(k,\Delta=-\Tilde{\Delta})\mathcal{T_+}^{-1}&=-h(-k,\Delta=-\Tilde{\Delta});\quad\mathcal{T_-}\mathcal{T_-^*}=\pm1,
\end{align*}
where $\mathcal{C_-}=\mathcal{T_+}=\sigma_x$.
Based on these conditions, we can conclude that the model possesses PHS ($\textrm{PHS}^{\dagger}$) when $\Delta=\Tilde{\Delta}$ ($\Delta=-\Tilde{\Delta}$), and fall in the $\pmb{\mathrm{BDI}}$ class ($\pmb{\mathrm{D}^{\dagger}}$) in $\pmb{\mathrm{AZ}}$ ($\pmb{\mathrm{AZ}^{\dagger}}$) symmetry classification.

\begin{figure}[ht]
    \includegraphics[width=0.5\textwidth, height=0.65\columnwidth]{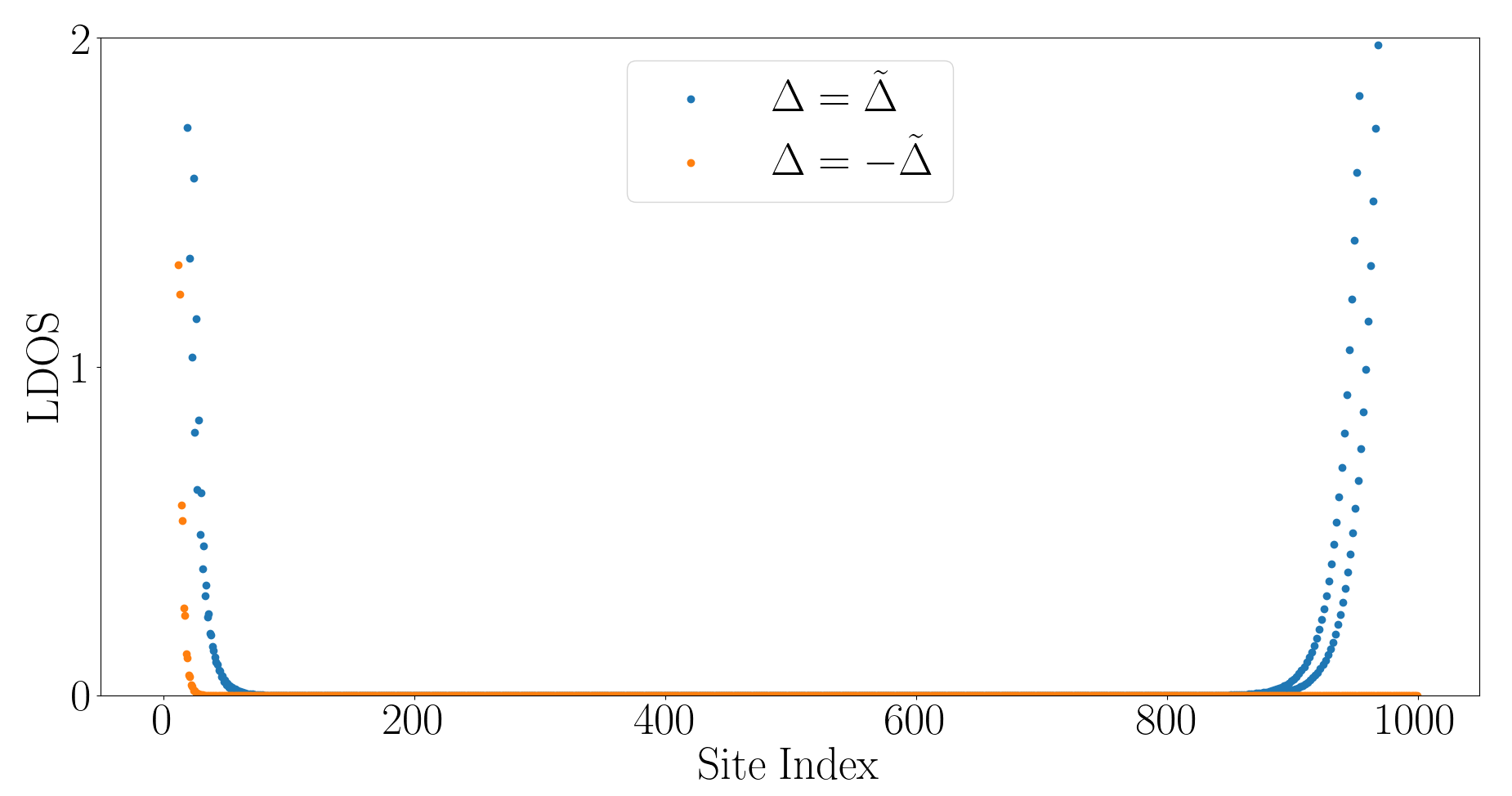}
    \caption{NHSE at left or both edge(s) for the cases $\Delta=\pm\Tilde{\Delta}$, respectively. The LDOS for both cases vanishes inside the chain and has finite value only at the edges.}
    \label{IPR_nr_2}
\end{figure}
We now investigate the localization phenomena by computing the LDOS.
As depicted in figure \ref{IPR_nr_2}, the LDOS exhibits pronounced accumulation either at one edge or both the edges of the 1D chain, contingent upon the specific relationship between $\Delta$ and $\Tilde{\Delta}$.
This breakdown of the BBBC unequivocally indicates the presence of NHSE in both scenarios.
For $\Delta=-\Tilde{\Delta}$, represented by orange triangles in figure \ref{IPR_nr_2}, the LDOS appears to have a higher value at the left edge, indicating that a majority of the bulk wave functions are localized at this edge.
This observation elucidates the usual NHSE, with the accumulation of the states at the left edge.

In contrast, when $\Delta=\Tilde{\Delta}$, there are higher values of LDOS at both edges, suggesting that most of the bulk eigenstates are localized at both edges simultaneously.
As demonstrated in Ref. \cite{PhysRevB.108.L060204}, PHS compels two `\textit{particle-hole}' partner skin modes to localize at opposite boundaries.
These skin modes, although localized at different boundaries and connected by PHS, are not degenerate, and instead exhibit opposite eigenvalues.
Our non-$\mathcal{PT}$ symmetric NH model having TRS ($\Delta=\Tilde{\Delta}$), also adheres to PHS.
This particular symmetry leads to non-degenerate skin modes localized at both the edges, which are related by the PHS.
When $\Tilde{\Delta}$ changes sign, the system forfeits PHS but acquires $\textrm{PHS}^{\dagger}$, resulting in NHSE to occur exclusively on one edge (left edge here).
It is noteworthy to mention that this phenomenon is very similar to the bi-directional NHSE \cite{PhysRevB.109.L081108}, which requires the system to have $\textrm{TRS}^{\dagger}$.
However, unlike our case, the skin modes are degenerate and localized at both the edges.

It is intriguing how a simple change in the sign of the non-reciprocity parameter, $\Tilde{\Delta}$, can drive the system from normal NHSE to NHSE at both the edges.
This demonstrates the system's sensitivity to parameter variations and offers promising avenues for controlling and manipulating NH phenomena for various applications.
As this particular system does not have BBBC, we refrain ourselves from further calculations involving the $k$-space.

\section{\label{sec4}$\mathcal{PT}$ symmetric NH model}

Now we break the Hermiticity of $H_1$ by including a non-reciprocity parameter, $\delta$, in the hopping term ($t_{ps}$) among the $s$ and $p_x$ orbitals from neighbouring unit cells.
The new Hamiltonian in OBC takes the form,
\begin{align}
		H_4=H_1+\delta\left[\sum_{i=1}^{L-1}(\hat{s}_{i}^{\dagger}\hat{p}_{i+1}-\hat{p}_{i+1}^{\dagger}\hat{s}_{i})-\sum_{i=2}^{L}(\hat{s}_{i}^{\dagger}\hat{p}_{i-1}-\hat{p}_{i-1}^{\dagger}\hat{s}_{i})\right]
    \label{eq:Ham4}
\end{align}
with the Bloch Hamiltonian, $h_4(k)$, given by,
\begin{equation}
		h_4(k)=\begin{pmatrix}
			\epsilon-2t\cos k & 2i(t_{ps}+\delta)\sin k\\-2i(t_{ps}-\delta)\sin k & -\epsilon+2t\cos k
		\end{pmatrix}.
    \label{eq:kspace_4}
\end{equation}
which indicates that the forward ($s^i\rightarrow p_x^{i+1}$) and the backward ($s^i\leftarrow p_x^{i+1}$) hopping amplitudes between $s^i$ and $p_x^{i+1}$ are $t_{ps}-\delta$ and $t_{ps}+\delta$, respectively.
The non-reciprocity term does not affect the TRS (satisfies equation \eqref{eq:TRS}) or the PHS (relevant to this case), as $h_4(k)$ satisfies,
\begin{equation*}
		\mathcal{C}_-h_4^T(k)\mathcal{C}_-^{-1}=-h_4(-k);\quad\mathrm{with}\quad\mathcal{C}_-\mathcal{C}_-^*=\pm1,
\end{equation*}
where $\mathcal{C}_-$ is unitary matrix and is equal to $\sigma_x$.
Hence, $h_4(k)$ has CS but does not have the sublattice symmetry (SLS), as it is defined via,
\begin{equation*}
		\mathcal{S}h(k)\mathcal{S}^{-1}=-h(k);\quad\mathrm{with}\quad\mathcal{S}^2=1,
\end{equation*}
for any $h(k)$, where $\mathcal{S}$ denotes a unitary matrix.
While CS and SLS are identical for a Hermitian system, they differ for NH systems.
For NH systems, the CS operator, $\Gamma$ demands the following relation to hold,
\begin{equation*}
    \Gamma h^{\dagger}(k)\Gamma^{-1}=-h(k);\quad\mathrm{with}\quad\Gamma^2=1,
\end{equation*}
for any arbitrary NH Bloch Hamiltonian, $h(k)$, it is evident that $h(k)\neq h^{\dagger}(k)$.
Thus, the system falls in class $\pmb{\mathrm{BDI}}$ in real $\pmb{\mathrm{AZ}}$ symmetry class due to the presence of all the three symmetries, namely, TRS, PHS, and CS.
Apart from this, the model also has $\mathcal{PT}$ symmetry, which will be shown later.

\begin{figure}[h]
	\includegraphics[width=0.5\textwidth, height=0.6\columnwidth]{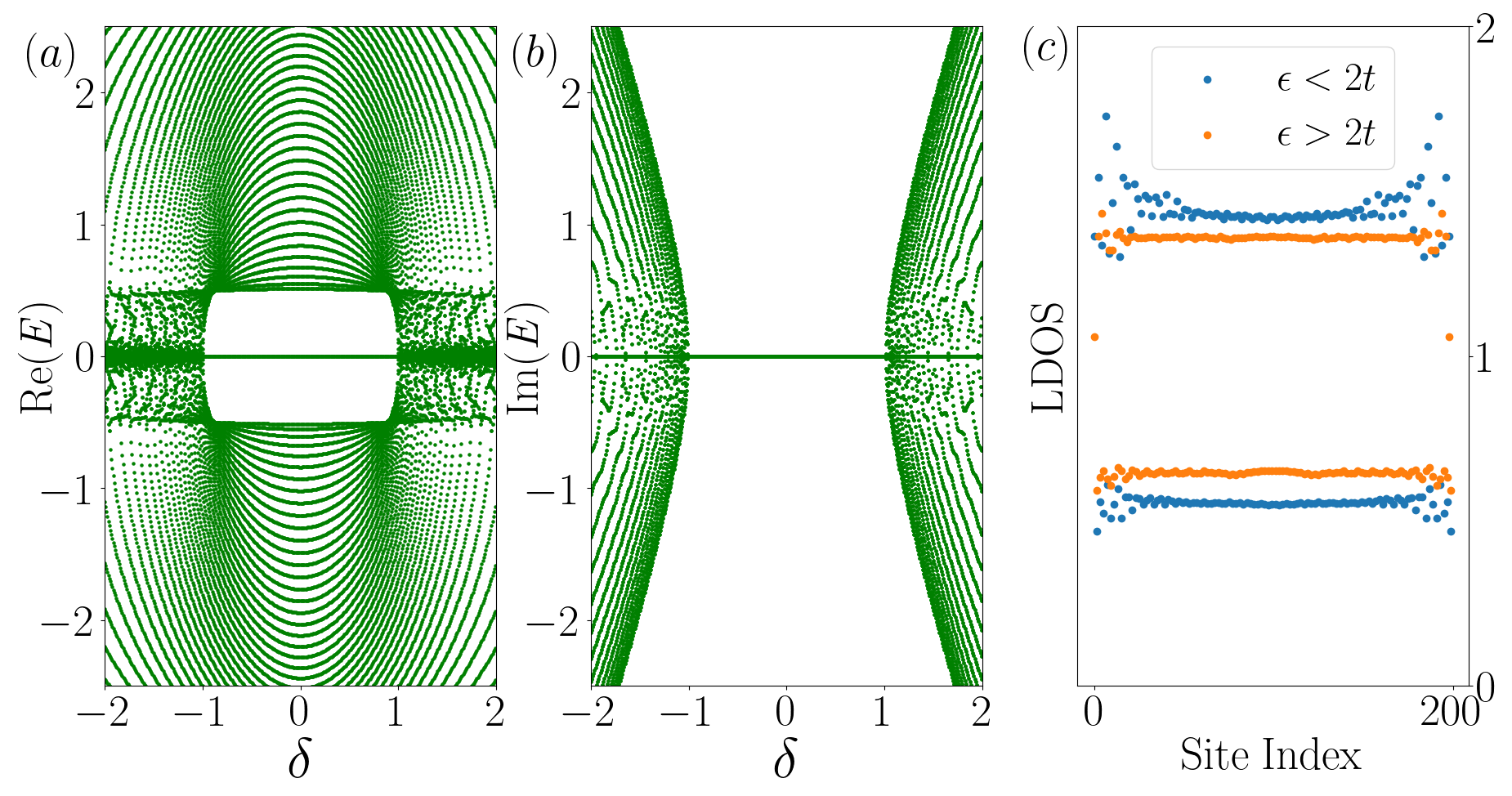}
		\caption{(Color online) The (a) real part, (b) imaginary part of eigenspectra are plotted as a function of the non-reciprocity parameter, $\delta$, with $\epsilon=1.5t$, $t=1,\;t_{ps}=1.5$. (c) The LDOS is calculated for $\epsilon=1.5$ ($<2t$) and $\epsilon=3$ ($>2t$). The LDOS confirms the absence of NHSE.}
    \label{rs_3}
\end{figure}

We shall explore the evolution of properties by varying the parameters, $\epsilon$ (onsite potential) and $\delta$ (non-reciprocity).
The eigenspectra in real space for the Hamiltonian, given by equation \eqref{eq:Ham4}, is plotted in figure \ref{rs_3} for $\epsilon<2t$.
Figures \ref{rs_3}(a) and \ref{rs_3}(b) show the real part (Re($E$)) and the imaginary part (Im($E$)) of the eigenvalues.
The plot indicates that the energies appear as ($E, E^*$) and ($E, -E$) pairs due to TRS and PHS, respectively.
Im($E$) is zero as long as the condition $|\delta|\leq t_{ps}$ is satisfied, suggesting that this is the $\mathcal{PT}$-unbroken phase \cite{PhysRevLett.80.5243}.
Beyond which, that is, for $|\delta|>t_{ps}$, Im($E$) becomes non-zero, indicating that the system has made an entry to a $\mathcal{PT}$-broken phase.
Figure \ref{rs_3}(c) represents the LDOS plots via two pairs of (almost) flat \emph{`lines'} corresponding to $\epsilon<2t$ (lines in blue) and $\epsilon>2t$ (orange lines).
These lines denote discrete values of LDOS at the lattice sites and oscillate from one site to the next.
This observation indicates the absence of NHSE.
Further, for $\epsilon<2t$, the blue lines demonstrate the existence of edge modes via higher values at the edges, while the orange ones do not have any such signature.
\begin{figure}[h]
		\includegraphics[width=0.5\textwidth, height=0.65\columnwidth]{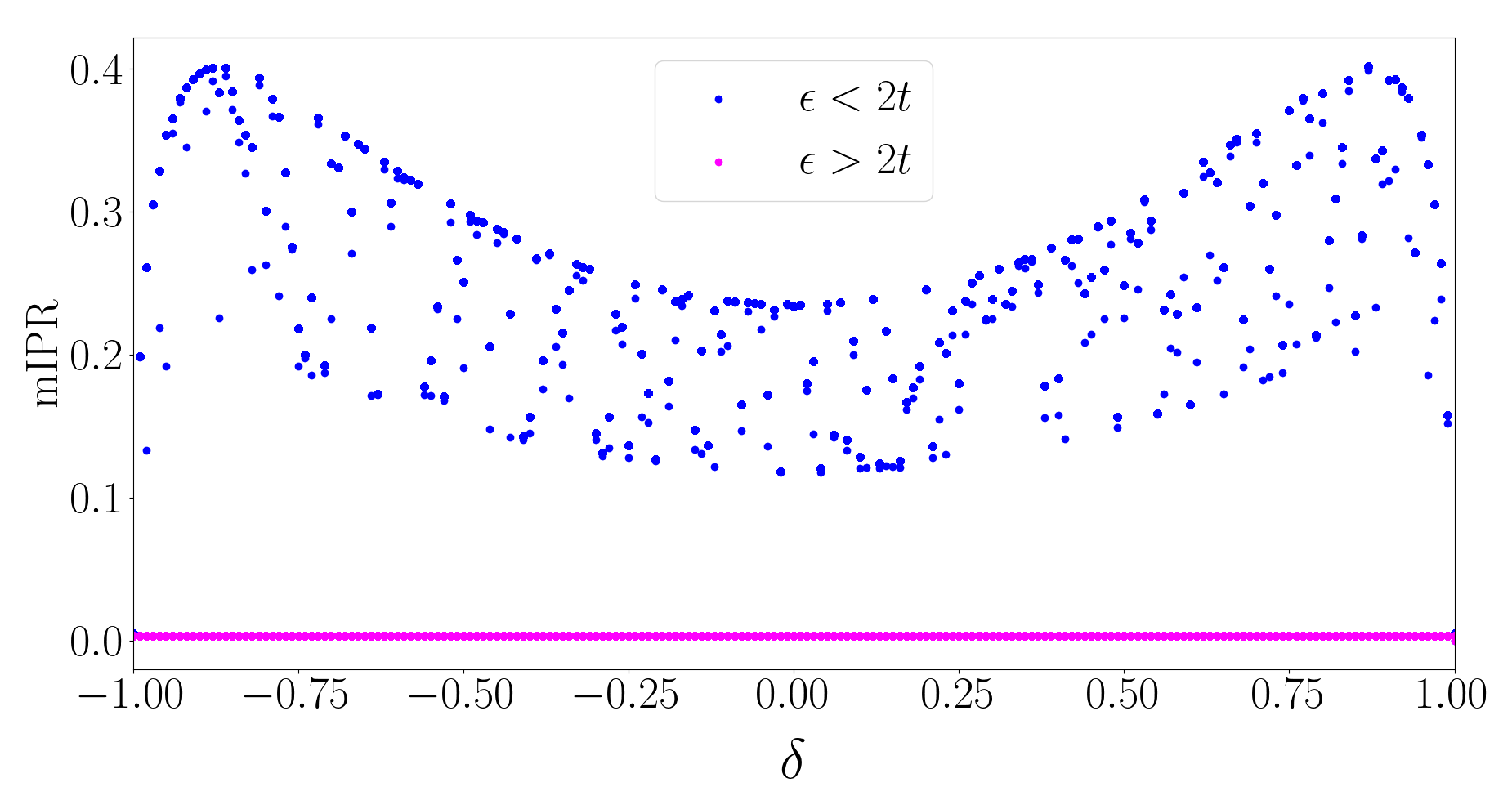}
		\caption{(Color online) The mIPR is varied with $\delta$ for $t=t_{ps}=1$ corresponding to the cases $\epsilon=1.2$ ($<2t$) and $\epsilon=2.6$ ($>2t$), respectively.}
    \label{ipr2}
\end{figure}

The degree of localization of these edge states, measured by mIPR, is shown in figure \ref{ipr2} as a function of the non-reciprocity parameter, $\delta$, in the range [$-1:1$].
The mIPR is non-zero for $\epsilon<2t$ and supports the existence of the edge states, suggesting that this is a topological phase.
These edge modes vanish as soon as $\epsilon$ becomes larger than $2t$ when all the eigenstates become extended, and mIPR vanishes.
Further, the system enters into a trivial phase, and thus a phase transition occurs at $\epsilon=2t$.

Let us analyze the reason behind the absence of NHSE in this NH model.
NH Hamiltonians with real eigenvalues necessarily belong to a particular class, namely the pseudo-Hermitian Hamiltonians \cite{10.1063/1.1418246}.
The condition for pseudo-Hermiticity for an arbitrary NH Hamiltonian, $H_{NH}$, can be stated as,
\begin{equation*}
     H^{\dagger}_{NH}=\eta\;H_{NH}\;\eta^{-1}
\end{equation*}
where $\eta$ is the pseudo-Hermitian operator and $H_{NH}$ is called $\eta$-pseudo-Hermitian Hamiltonian.
Specifically, consider an operator, $\Tilde{\eta}$, for which $h_4(k)$ (equation \eqref{eq:kspace_4}) also satisfies the following relation,
    \begin{equation}
        h^{\dagger}_4(k)=-\Tilde{\eta}\;h_4(k)\Tilde{\eta}^{-1},
        \label{eq:pshh}
    \end{equation}
where $\Tilde{\eta}=U^{\dagger}\sigma_zU$ with $U=\frac{1}{\sqrt{2}}\begin{pmatrix}
        1 & -1\\
        1 & 1
\end{pmatrix}$ and $\sigma_z$ is $z$-component of the Pauli matrix.
Hence, $h_4(k)$ qualifies as a $\Tilde{\eta}$-pseudo-skew-Hermitian Hamiltonian, since it anticommutes with $\Tilde{\eta}$.
By this argument, the system (obeying equation \eqref{eq:pshh}) exhibits the behavior of a Hermitian system (owing to the presence of pseudo-skew-Hermiticity).
Since NHSE is not a feature of Hermitian systems, it does not occur in our case.
It is crucial to emphasize that the safeguarding against NHSE attributed to $\Tilde{\eta}$-pseudo-skew-Hermiticity is not universally applicable.
Recent studies \cite{PhysRevB.105.245407, PhysRevB.109.165407} have demonstrated the persistence of NHSE even in systems characterized by pseudo-Hermiticity. Thus, the absence of NHSE is specific to this model.

Now, we shall analyze the system in $k$-space using the Bloch Hamiltonian given in equation \eqref{eq:kspace_4} to establish a correspondence with the real space.
We shall proceed in the same manner as in the previous sections, that is, the basis of the Bloch Hamiltonian will be `rotated' via a unitary operator, $U$, given in equation \eqref{eq:U}.
The resultant Bloch Hamiltonian, let us call it as $h'_4(k)$, is given by,
\begin{gather}
		h'_4(k)=\begin{pmatrix}
			2i\delta\sin k & -\epsilon+2t\cos k\\ &+2it_{ps}\sin k\\-\epsilon+2t\cos k\\-2it_{ps}\sin k &-2i\delta\sin k
		\end{pmatrix},
    \label{eq:kspace_7}
\end{gather}
with $\pmb{d'_4}\equiv(-\epsilon+2t\cos k, -2t_{ps}\sin k, 2i\delta\sin k)$, suggesting that all, but the $z$ component of $\pmb{d'_4}$ is real, that is,
\begin{align*}
		d_{4x}^{'\mathrm{R}}&=-\epsilon+2t\cos k,\;d_{4y}^{'\mathrm{R}}=-2t_{ps}\sin k,\;d_{4z}^{'\mathrm{R}}=0;
		\\d_{4x}^{'\mathrm{I}}&=0,\;d_{4y}^{'\mathrm{I}}=0,\;d_{4z}^{'\mathrm{I}}=2i\delta\sin k,
\end{align*}
where the notations bear a similar meaning to those given in the previous section.
It is clear that $h'_4(k)$, which is connected to $h_4(k)$ via a similarity transformation, commutes with the $\mathcal{PT}$ operator given by $\sigma_x\mathcal{K}$ and hence preserves $\mathcal{PT}$ symmetry.
Note that, equation \eqref{eq:pshh} illustrates a more general aspect of the $\mathcal{PT}$ symmetry inherent in $h_4'(k)$.
The expression for the band structure is given by,
\begin{align}
		E_{4\pm}(k)&=\pm\sqrt{(d_{4x}^{'\mathrm{R}}+id_{4x}^{'\mathrm{I}})^2+(d_{4y}^{'\mathrm{R}}+id_{4y}^{'\mathrm{I}})^2+(d_{4z}^{'\mathrm{R}}+id_{4z}^{'\mathrm{I}})^2}\nonumber\\
		&=\pm\sqrt{(-\epsilon+2t\cos k)^2+4(t_{ps}^2-\delta^2)\sin^2 k}.
		\label{bs_3}
\end{align}
Moreover, the left and the right eigenvectors of $h'_4(k)$ are obtained as,
\begin{align}
		\ket{\lambda'_{4\pm}(k)}&=\pm\frac{1}{\sqrt{2}}\beta_1e^{\pm i\alpha^*}
		\begin{pmatrix}
			\frac{\sin\theta^*_k e^{-i\phi_{4k}}}{\pm 1+\cos\theta^*_k}\\1
		\end{pmatrix};\nonumber\\
		\ket{\psi'_{4\pm}(k)}&=\pm\frac{1}{\sqrt{2}}\beta_2e^{\pm i\alpha}
		\begin{pmatrix}
			\frac{\sin\theta_k e^{-i\phi_{4k}}}{\pm 1-\cos\theta_k}\\1
		\end{pmatrix},
    \label{eig_v_3}
\end{align}
where $\alpha$ is independent of $k$ and $\phi_{4k}$ and $\theta_k$ are given by,
\begin{align*}
		\phi_{4k}&=\tan^{-1}\left[\frac{-2t_{ps}\sin k}{-\epsilon+2t\cos k}\right];\\ \theta_k&=\tan^{-1}\left[\frac{\sqrt{(-\epsilon+2t\cos k)^2+4t_{ps}^2\sin^2k}}{2i\delta\sin k}\right].
\end{align*}
$\beta_1$ and $\beta_2$ satisfy the relation,
\begin{equation*}
		\beta^*_1\beta_2=\frac{1}{2}(1-\cos\theta_k)
\end{equation*}
and are periodic functions of $k$.
\begin{figure}[t]
	\includegraphics[width=0.5\textwidth, height=0.7\columnwidth]{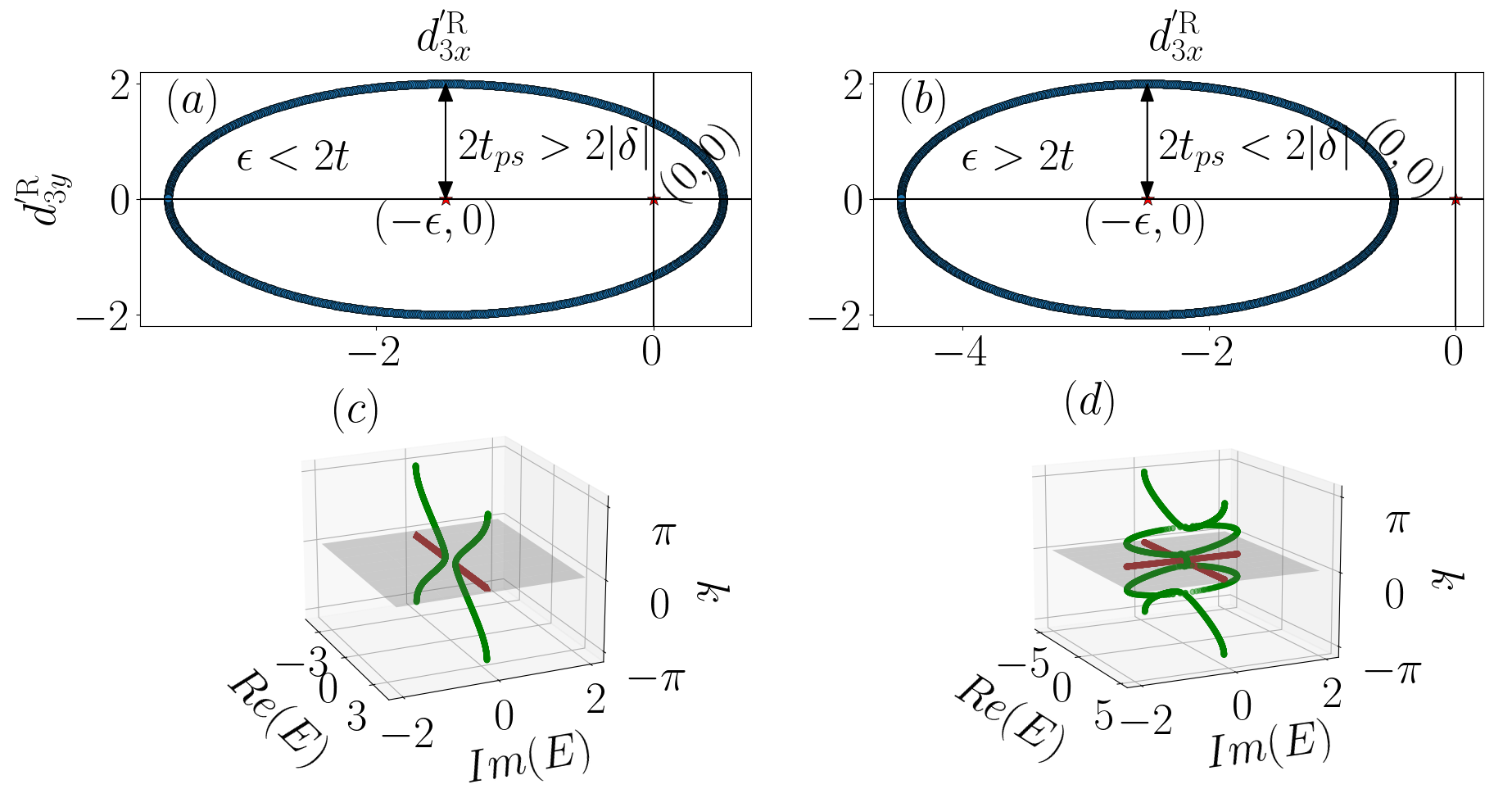}
		\caption{(Color online) The locus of the $\pmb{d'_4}$-vector is drawn in a plane spanned by $d_{4x}^{'\mathrm{R}}-d_{4y}^{'\mathrm{R}}$ for the parameters $t=1$, $t_{ps}=1$ and (a) $\epsilon=1.5$, where the ellipse contains the origin ($0, 0$) (topological case) and (b) $\epsilon=2.5$, where it excludes the origin (trivial case). (c) and (d) represent 3D figures of the band structures ($E_{4\pm}(k)$ vs $k$) with the real and the imaginary parts of the energy and the momentum ($k$) being plotted along the $x$, $y$, and $z$ directions, respectively. The parameters are the same as those of (a) and (b). The points in green and brown are the actual data points and their 2D projections, respectively.}
    \label{band3}
\end{figure}
\begin{figure}[b]
		\includegraphics[width=0.5\textwidth, height=0.6\columnwidth]{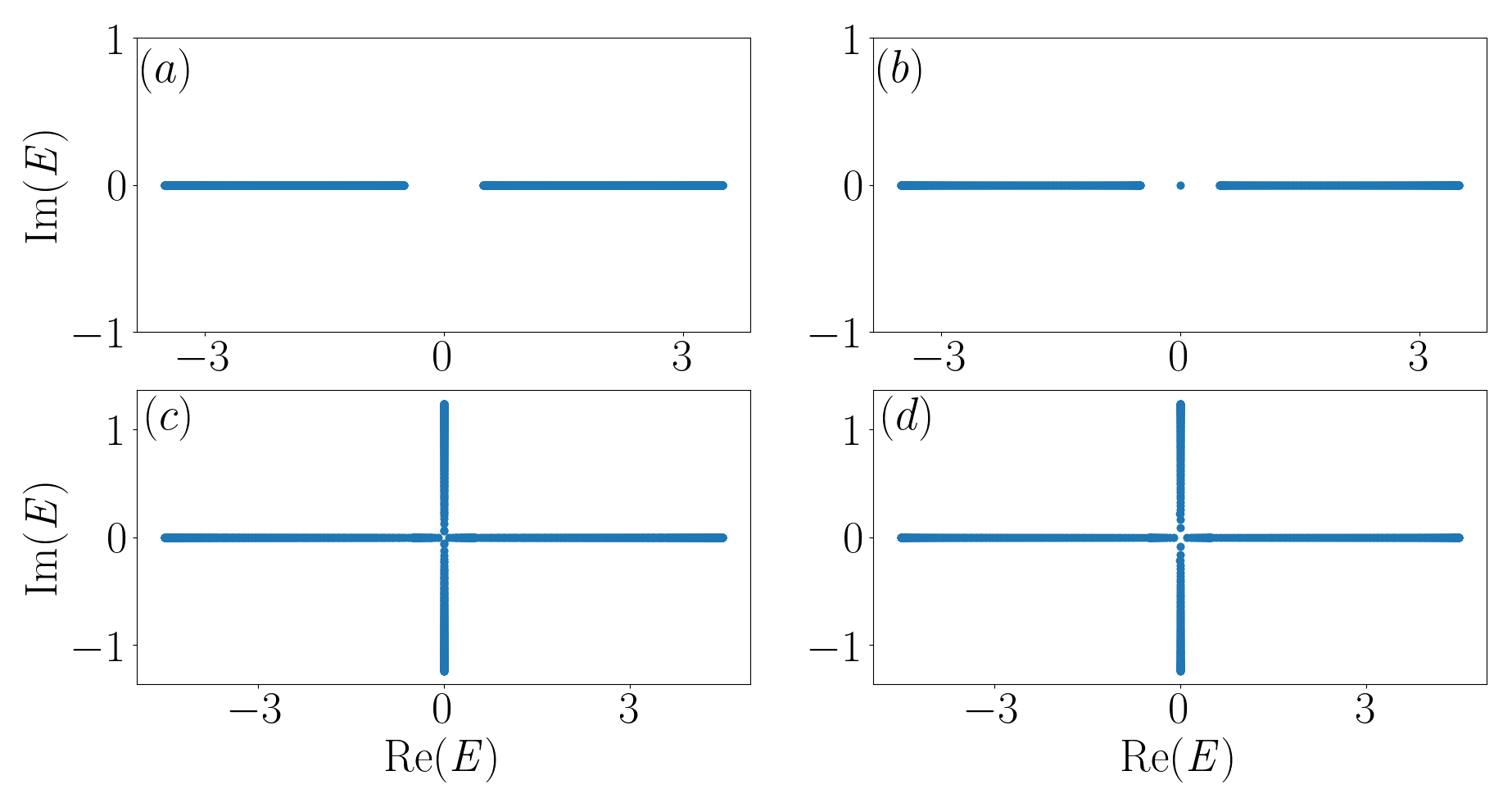}
		\caption{(Color online) Real and imaginary parts of energy, corresponding to the Hamiltonian $H_4$, for PBC (left panel) and OBC (right panel). The parameters are kept as follows: $\epsilon<2t$ for both (a) and (b), while $\epsilon>2t$ for both (c) and (d).}
		\label{rm_n}
\end{figure}
The energy $E_{4\pm}(k)$ in equation \eqref{bs_3} becomes zero when $d_{4x}^{'\mathrm{R}}=0$ and $d_{4y}^{'\mathrm{R}}=|d_{4z}^{'\mathrm{I}}|$.
These two are the conditions for an EP.
Putting these conditions in equation \eqref{eig_v_3}, we see that, $$\left<\lambda'_{4\pm}(k)|\psi'_{4\pm}(k)\right>=0.$$
This tells us that the bi-orthogonality condition gets violated at the EP.

A similar picture emerges for figures \ref{band3}(a) and \ref{band3}(b), where the trajectory of $\pmb{d'_4}$, which is an ellipse, is represented in a plane spanned by $d_{4x}^{'\mathrm{R}}-d_{4y}^{'\mathrm{R}}$ for two values of $\epsilon$, namely, $\epsilon=1.5$ and $\epsilon=2.5$, respectively.
In figure \ref{band3}(a), the origin ($0,0$) is inside the ellipse, that is, $\epsilon<2t$, suggesting that it denotes a topological phase.
On the other hand, figure \ref{band3}(b) describes the trivial phase as the ellipse excludes the origin when $\epsilon>2t$.
Similar to the non-$\mathcal{PT}$ symmetric case (without TRS), the energy gap, in this case, too, represents a line gap.
Note that, the $\mathcal{PT}$-broken-unbroken phase transition has an analytic dependence on the parameters given by,
\begin{equation}
		\epsilon_0=2\sqrt{\delta^2+t^2-t_{ps}^2}.
    \label{eq:pt}
\end{equation}
Figure \ref{band3}(c) shows the case where $\mathcal{PT}$ symmetry is still unbroken since the condition $\epsilon>\epsilon_0$ is satisfied, resulting in the eigenspectra to be purely real (the imaginary part being zero shown by the brown line).
When the value of $\epsilon$ falls below the critical value, $\epsilon_0$, it corresponds to the $\mathcal{PT}$-broken phase.
This phase is depicted in figure \ref{band3}(d), where some of the eigenvalues become purely imaginary.
Finally, for large values of $|\delta|$, all the eigenvalues become imaginary and appear in pairs with an imaginary line gap.
Figure \ref{rm_n} demonstrates that the eigenvalue spectra exhibit insensitivity to the boundary conditions, as they remain nearly identical for the PBC and the OBC in the topological and trivial regions.
Adhering to the properties of pseudo-Hermitian Hamiltonians outlined in Ref. \cite{10.1063/1.1418246}, it can be demonstrated that, for pseudo-skew-Hermitian Hamiltonians, one of the following conditions must hold,
\begin{enumerate}
    \item The eigenvalues of the Hamiltonian are real and come in positive and negative pairs, that is, $(\pm E)$.
    \item The complex eigenvalues come in negative complex conjugate pairs with opposite signs, that is, $(E, -E^*)$.
\end{enumerate}
Figure \ref{rm_n} provides a clear visualization of the above conditions.

Putting $\ket{\lambda'_{4\pm}(k)}$ and $\ket{\psi'_{4\pm}(k)}$ from equation \eqref{eig_v_3} into equation \eqref{Berry_NH}, we get an expression for the CBP as,
\begin{align}
		\gamma_{\pm}&=\frac{1}{2}\oint_{BZ}\frac{\partial\phi_{4k}}{\partial k}(1+\cos\theta_k)\;dk\nonumber\\=\frac{1}{2}&\oint_{BZ}\frac{2t_{ps}(\epsilon\cos k-2t)}{(-\epsilon+2t\cos k)^2+4t_{ps}^2\sin^2k}\left(1+\frac{2i\delta\sin k}{E_{4+}(k)}\right)dk
    \label{eq:berry3}
\end{align}
The phase diagram of CBP obtained from the equation \eqref{eq:berry3} is precisely similar to that of the figure \ref{B_phase} (and hence not shown again), hinting that the topological phase transition occurs at $\epsilon=2t$.

\begin{widetext}
For a lucid description, we present Table \ref{table:NHSE} to summarize the localization phenomena for different NH models, Which states the unique features of different NH models considered by us and enumerates the presence or absence of NHSE along with its variants.
\begin{table}[h]
\centering
\begin{tabular}{|c|c|c|c|c|}
        \hline
        Cases & \RNum{1} & \RNum{2} & \RNum{3} & \RNum{4}
        \\
        \hline
         NH & non-$\mathcal{PT}$ symmetric & non-$\mathcal{PT}$ symmetric  & non-$\mathcal{PT}$ symmetric & $\mathcal{PT}$ symmetric \\ models & model without & model with TRS & model with both & model \\ & TRS & but no PHS ($\Delta=-\Tilde{\Delta}$) & TRS \& PHS ($\Delta=\Tilde{\Delta}$) &
         \\
        \hline
        Cause of & onsite staggered & non-reciprocity in & non-reciprocity in & non-reciprocity in
        \\
        non-Hermiticity & imaginary potential & $t$ ($s\leftrightarrow s$ or $p_x\leftrightarrow p_x$) & $t$ ($s\leftrightarrow s$ or $p_x\leftrightarrow p_x$) & $t_{ps}$ ($s\leftrightarrow p_x$)
        \\
        \hline
        NHSE & no NHSE & normal NHSE & NHSE at both edges& no NHSE\\
        \hline
    \end{tabular}
    \caption{Presence or absence of NHSE in different NH models. $\Delta$ and $\Tilde{\Delta}$ are defined in equation \eqref{eq:Ham5}.\label{table:NHSE}}

\end{table}
\end{widetext}

\section{\label{sec5}Conclusion}
	
In this work, we have explored NH extensions of a one-dimensional spinless adaptation of the BHZ-like model, comparing it against its Hermitian counterpart to understand the effects of non-Hermiticity.
Our investigations into the localization behavior of the edge modes and topological characteristics of the NH extensions of the model reveal a range of fascinating phenomena.
Critical distinctions among various types of non-Hermiticity include their inherent symmetries, the pivotal parameters of the Hamiltonians influencing phase transitions, and the manifestation of NHSE.
The extension into the NH realm bifurcates the model into two distinct versions: a $\mathcal{PT}$ symmetric model and a non-$\mathcal{PT}$ symmetric model (with or without TRS).
Notably, we have found that in the $\mathcal{PT}$ symmetric system, a spontaneous $\mathcal{PT}$ symmetry breaking occurs, segregating the system into regimes with real or complex eigenvalues in the $\mathcal{PT}$-unbroken and $\mathcal{PT}$-broken phases, respectively.
Furthermore, the topological phases are succinctly explored via the complex Berry phase, serving as the topological invariant for NH systems obeying BBBC.
We integrate the findings with the emergence of localized zero-energy edge modes.

Hence with the above inputs, we have engineered a model wherein, depending on the choice of non-reciprocity, three different phenomena emerge, namely, a) No NHSE ($\mathcal{PT}$ symmetric model), b) normal NHSE (non-$\mathcal{PT}$ symmetric model with TRS), and c) NHSE at both edges (non-$\mathcal{PT}$ symmetric model with TRS and PHS).
Thus by simply tuning the parameters of the model, it is possible to realize the build-up of bulk states at either edge (normal NHSE), both edges, or even delocalization of them (no NHSE).
This may have crucial importance in the transfer of information over arbitrarily large distances.
This particular adaptability underscores the versatility of the model and the profound influence of non-reciprocal features on NH systems.
\bibliographystyle{vancouver}
\bibliography{ref2}
\end{document}